\documentclass[times,twocolumn,review]{elsarticle}

\usepackage{jasr}
\usepackage{framed,multirow}

\usepackage{amssymb}
\usepackage{latexsym}

\usepackage[switch]{lineno}

\usepackage{url}
\usepackage{xcolor}
\definecolor{newcolor}{rgb}{.8,.349,.1}
\usepackage{longtable}
\usepackage[citebordercolor=green]{hyperref}
\usepackage{amsmath}
\usepackage{lscape}
\usepackage{booktabs}

\newcommand{\kms}{   {km s$^{-1}$}}

\chardef\us=`\_

\journal{Advances in Space Research}

\begin{document}

\verso{Pooja Devi \textit{etal}}

\begin{frontmatter}

\title{Band Splitting in m-Type II radio Bursts and their Role in Coronal Parameter Diagnostics \tnoteref{tnote1}}%

\author[1,2,3]{Pooja \snm{Devi}\corref{cor1}}
\cortext[cor1]{Corresponding author:}
\ead{setiapooja.ps@gmail.com}
\author[1]{Ramesh \snm{Chandra}}
\author[4]{Rositsa \snm{Miteva}}
\author[5]{M. Syed \snm{Ibrahim}}
\author[1]{Kamal \snm{Joshi}}

\address[1]{Department of Physics, DSB Campus, Kumaun University, Nainital-263001, India}
\address[2]{Rosseland Centre for Solar Physics, University of Oslo, P.O. Box 1029, Blindern, N-0315 Oslo, Norway}
\address[3]{Institute of Theoretical Astrophysics, University of Oslo, P.O. Box 1029, Blindern, N-0315 Oslo, Norway}
\address[4]{Institute of Astronomy and National Astronomical Observatory - Bulgarian Academy of Sciences, 72 Tsarigradsko Chaussee Blvd., Sofia 1784, Bulgaria}
\address[5]{Department of Physics, Sri Sai Ram Engineering College, Sai Leo Nagar, West Tambaram, Chennai – 600 044 Tamil Nadu, India.}

\communicated{P. Devi}

\begin{abstract} 
Type II radio bursts are signatures of shock waves generated by solar eruptions, observed at radio wavelengths. While metric (m) type II bursts originate in the lower corona, their longer-wavelength (up to kilometers) counterparts extend into interplanetary space. A rare but valuable feature observed in some type II bursts is band splitting in their dynamic spectra, which provides crucial insights into physical parameters such as shock speed, Alfv\'en Mach number, Alfv\'en speed, and coronal magnetic field strength (B). In this study, we investigate band-splitting in 44 m-type II radio bursts observed by the Radio Solar Telescope Network during solar cycle 24 (2009 -- 2019). These events exhibit splitting in both fundamental and harmonic bands and are analyzed under both perpendicular and parallel shocks. All events are associated to solar flares and 41 (93 \%) with the coronal mass ejections. Shock speeds, derived using a hybrid coronal density model proposed by \cite{Vrsnak2004}, range from $\approx$ 350 to 1727 \kms. The relative bandwidth (BDW) of the split bands remains constant with frequency and height. Alfv\'en Mach numbers indicate moderate shock strength (1.06 -- 3.38), while Alfv\'en speeds and $B$ vary from $\approx$ 230 -- 1294 \kms\ and $\approx$ 0.48 -- 7.13 G, respectively. Power-law relationships are established as $BDW \propto f_L^{-0.4}$ and $BDW \propto R^{\sim1}$, while the coronal magnetic field decreases with height as $B \propto R^{\sim-3}$. These results enhance our understanding of shock dynamics and magnetic field structures in the solar corona.

\end{abstract}


\end{frontmatter}


\section{Introduction}
\label{sec:introduction}

Observations of radio emission from the Sun are most frequently organized in a frequency--time plot, where the intensity was initially ordered as a profile view \citep{1960BAN....15..229D} but has recently been color-coded. This particular type of plot is termed the dynamic radio spectrum. Its advantages are that low-cost instrumentation could supply high-frequency and high-temporal resolution data, and a network of similar stations could provide a nearly complete universal time (UT) coverage. The most important disadvantage is the position loss of the radio emission on the solar disk. However, the shapes of the radio features in the dynamic spectrum (historically enumerated from type I to V) point towards the identity and direction of the driver \citep{2004LNP...656...49W}. Furthermore, when coupled with a density model, the height over the solar surface of the emitting source could be deduced \citep{Lawrance2024}. Therefore, radio bursts carry rich potential for coronal diagnostics. In this study, we focus on the so-called type II bursts.

Solar Type II radio bursts are radio emissions characterized by a slow frequency drift, observable in solar dynamic spectra. The drift rates (rate of change of frequency with time) range from 0.1 to 1 MHz s$^{-1}$ in the metric (m) domain (upper corona), and decrease to below 0.01 MHz s$^{-1}$ in the decametric–hectometric (DH) and kilometric domains (interplanetary (IP) space). The m-type II bursts are discovered by \citet{wild1950} and have been reported by different research groups \cite[e.g.,][]{Ginzburg1958, Thejappa2000, Gopalswamy2012, Frassati2019, 2021SoPh..296...27U, Lawrance2024, Devi2024}. These bursts often exhibit two distinct components: the fundamental and the harmonic. These correspond to radio emission at the local plasma frequency and twice the plasma frequency, respectively. For a long time, it was believed that type II radio bursts are the signatures of magnetohydrodynamic (MHD) shock waves \citep{Nelson1985} and can be generated by coronal mass ejections (CMEs), solar flares, and fast plasma flows in regions of reconnection of magnetic field lines \citep{Mann1995}. Further, the onset frequency of Type II radio bursts can serve as a diagnostic for estimating the coronal density at which the CME-driven shock originates.
One of the key mechanisms behind the generation of type II solar radio bursts is the Buneman instability \citep{Pikelner1963, Fomichev1972, Zaitsev1977}. This phenomenon arises when a fast-moving shock wave travels through the solar plasma, inducing strong 
currents. These currents, in turn, lead to the formation of MHD shocks or collective plasma waves in the solar corona and IP space, leading to the generation of observed radio emissions. Recently, \cite{Chernov2021} completed a study on the origin of type II radio bursts. The authors proposed a model to explain these bursts under the framework of the Buneman instability. According to their model, this is fulfilled when two conditions are met: (1) the Alfv\'en Mach number is higher than a certain limit, and (2) the shock front is nearly perpendicular.
 
Occasionally, these type II radio bursts show splitting in two parallel emission bands for a single type II burst in the dynamic spectrum. These splitting can be observed in both the fundamental and harmonic components \citep{Smerd1974,Smerd1975}. 
In order to explain observed type II band splitting, \cite{Smerd1974} proposed a model according to which the band spitting arises from coherent plasma emission produced in both the upstream and downstream regions of a propagating ahead and behind of the coronal shock wave, respectively. According to this interpretation, the frequency separation between the bands corresponds to the density jump across the shock, with the upstream emission corresponding to the lower-density plasma ahead of the shock and the downstream emission to the denser, compressed plasma behind it. The upstream and downstream are generally denoted by lower frequency branch (LFB) and higher frequency branch (HFB), respectively.

Previously, \citet{Mann1996} studied the 25 m-type II radio bursts observed by Institute for Astrophysics Potsdam spectrograph in 40-170 MHz and calculated the relative instantaneous bandwidth. According to them {\it ``the instantaneous bandwidth of a solar type II radio bursts would reflect the density jump across the associated shock wave''}. Later on, \citet{Vrsnak2001} studied the band spitting in coronal and IP type II radio bursts and computed the basic parameters such as: relative band splitting (BDW), and frequency drift, which are important parameters for the estimation of the shock characteristics. Further, \citet{Vrsnak2002} and \citep{Vrsnak2004}
extended their study of band spitting over 18 m-type II bursts and computed the Alfv\'en velocity and the coronal magnetic field as a function of the heliocentric distance.

Thus, the most commonly used model to explain the band splitting is by \cite{Smerd1974}, as UFB and LFB can be directly used to compute different coronal parameters such as density, Alfv\'en Mach number, and magnetic field strength \citep{Vrsnak2002,Mahrous2018,Zucca2018}. After the availability of imaging observations of type II radio bursts this scenario was supported by \cite{2012A&A...547A...6Z,Zucca2014, 2018ApJ...868...79C}, for selected case studies. On the other hand, alternative models exist to explain the band-splitting due to two different radio sources at two different locations in the upstream of the shock front \citep{Zaitsev1978,Holman1983}. Recently, \cite{Normo2025} studied the event of band splitting with LOFAR data on 23 May 2022 and found that both bands originate in two separate upstream regions. Despite the alternative theories, we decided to adopt the one by \cite{Smerd1974} in our study mostly due to its straight-forward deduction of physical parameters and the lack of imaging data for our entire event list.

In this study we adopt the premise that the frequency separation between the split bands provides an estimate of the density compression ratio across the shock. This allows inferring the shock parameters such as: the Mach number and finally the coronal magnetic field strength, which are difficult to measure directly.
Since the band spitting is more clearly visible in the m-region ($\sim$300 MHz frequency band), in this paper, we present the m-type II radio bursts with band splitting and their relationship with various flare, CME, and other coronal parameters (such as Alfv\'en Mach number, height of shock formation, and coronal magnetic field strength). Section \ref{sec:data} describes in details the data sets and the selection criteria. The subsequent parts are devoted to the obtained results (Section \ref{sec:results}) and their discussion (Section \ref{sec:discussion}). Finally, the summary of the study is structured in Section \ref{sec:summary}. 

\section{Data Sets}
\label{sec:data}
\begin{figure}[t]  
    \centering
    \includegraphics[width=0.485\textwidth]{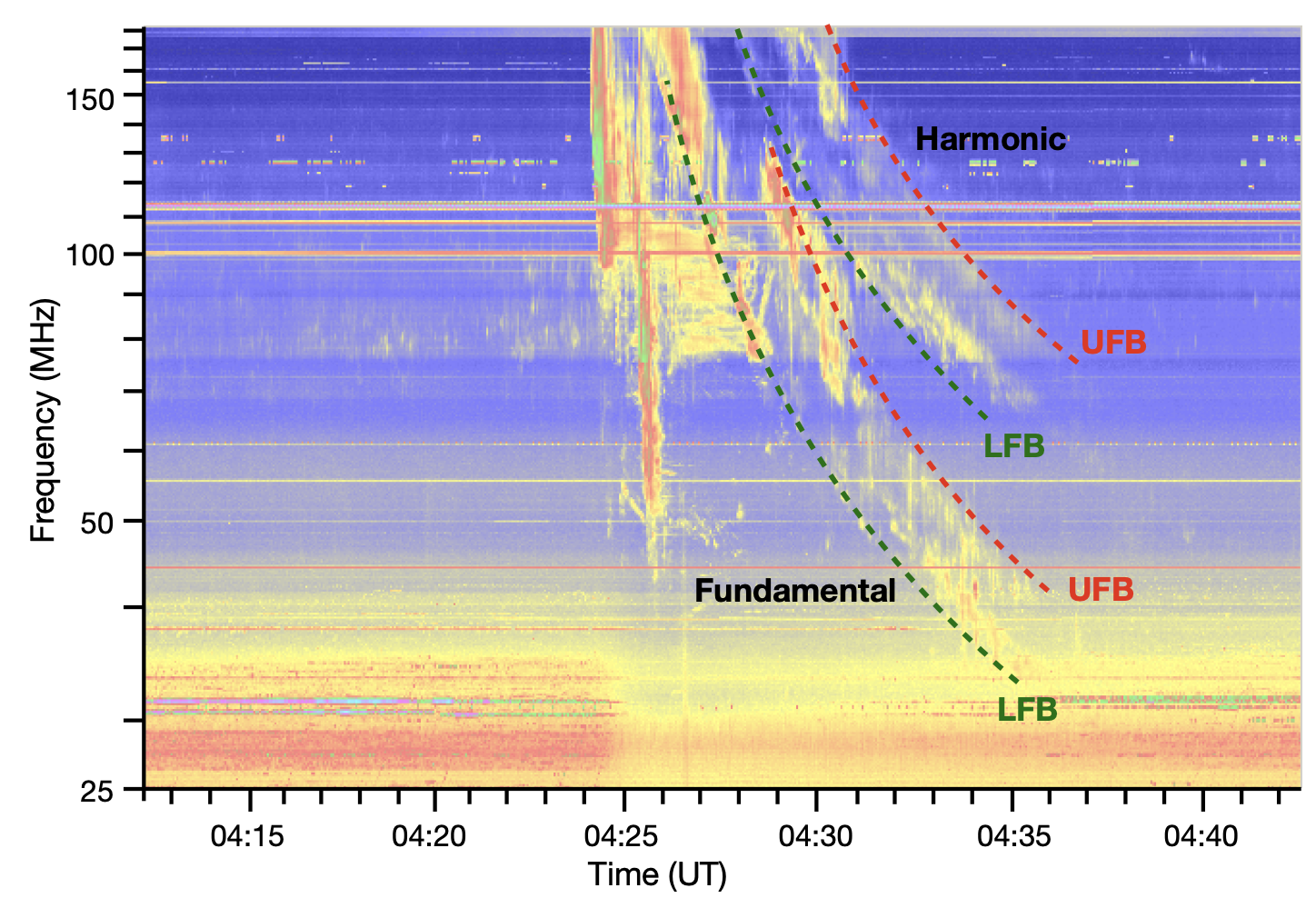}
    \caption{An example of m-type II radio burst from Learmonth Hill on 08 November 2013 showing the fundamental and harmonic bands. The UFBs and LFBs are shown with red and green dashed curves, respectively. }
    \label{fig:example}
\end{figure}

In this study, we used data from the Radio Solar Telescope Network (RSTN) which covers the radio spectrum in the frequency range 25 -- 180 MHz with its four stations, namely Learmonth Hill, San Vito, Palehua, and Sagamore Hill, covering the full day observations.  
We started from the type II radio burst catalog as compiled by \cite{Lawrance2024} and \cite{Devi2024}. From their list, we selected the events that show clear band splitting in both fundamental and harmonic bands. Note that the frequency ratio of these splitted bands are different from $\frac{f_{harmoic}}{f_{fundamental}}$ = 2 (nearly) as defined in \citet{Vrsnak2001}, which means that they are different from what are called as ``fundamental'' and ``harmonic'' bands. 
Of the 429 m-type II radio bursts observed by RSTN during solar cycle 24 in the list of \cite{Lawrance2024}, we found clear band splitting in 44 events. 
An example of band splitting in the fundamental and harmonic bands on 08 November 2013 from the Learmonth Hill station is shown in Figure \ref{fig:example}. The figure clearly show the UFB and LFB in both fundamental and harmonic bands. 

After the selection of events, we checked their association with the solar flares, CMEs, and the DH type II radio bursts. The information about flares is taken from the NOAA solar region summary reports 
(\href{https://www.ngdc.noaa.gov/stp/space-weather/swpc-products/daily_reports/solar_event_reports/}{solar\_region\_summary\_reports}), CMEs from the LASCO CME catalog \citep[][\url{https://cdaw.gsfc.nasa.gov/CME_list/}]{Gopalswamy2009}, and of the DH type II bursts from the Wind/WAVES type II catalog \citep[][\url{https://cdaw.gsfc.nasa.gov/CME_list/radio/waves_type2.html}]{Gopalswamy2019}.
We found that all the events are associated with solar flares and 41 (93 \%) are associated with CMEs. Out of 44 m-type IIs, 12 events (i.e., $\approx$ 27 \%) are associated with the DH-type II radio bursts. Out of these 12 DH-type IIs, eight (i.e., $\approx$ 67 \%) are associated with halo CMEs.
A table containing the parameters of these band-splitted m-type IIs and their associated flares, CMEs, and DH type IIs, is presented in the Appendix as Table~\ref{table1}. The description about the associated flares, CMEs, and various parameters computed from these band-splitted m-type II bursts are provided in the following Section.

\section{Results}
\label{sec:results}

\subsection{Relationship with flares}
\label{sec:flare}
\begin{figure*}[!t] 
    \centering
    \includegraphics[width=\textwidth]{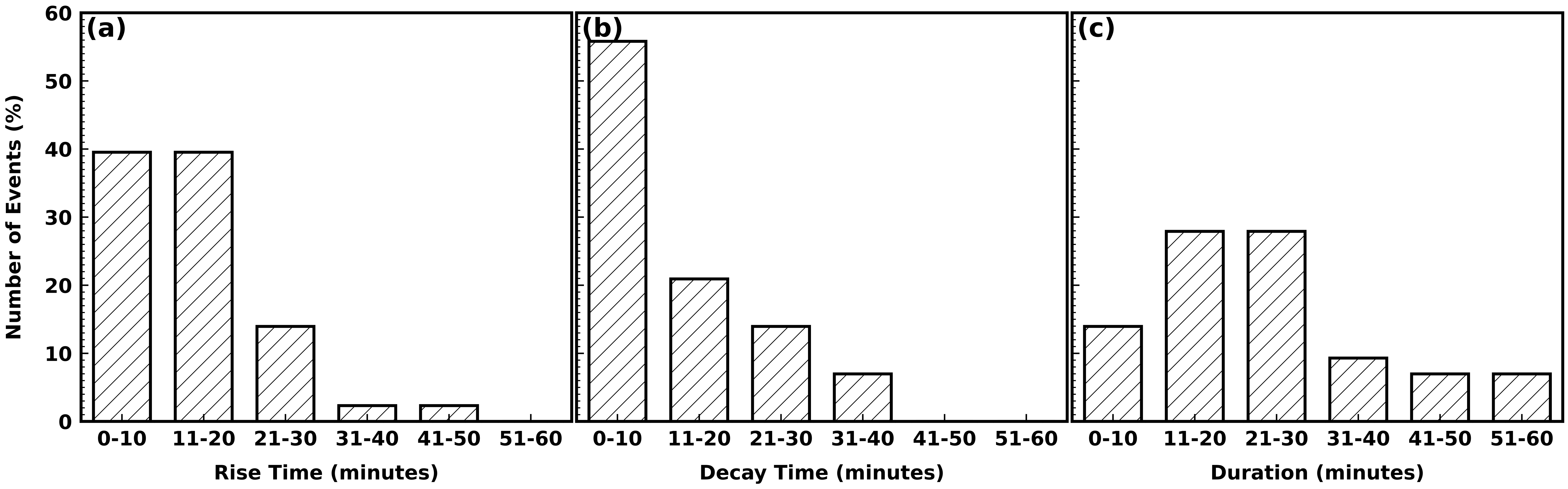}
    \caption{m-type II radio bursts associated flare parameters namely rise time, decay time, and duration of flares in panels (a), (b), and (c), respectively.}
    \label{fig:flare}
\end{figure*}

We cross-checked the relationship of the type IIs in this study with the flares and CMEs. We found that all of these type IIs are associated with moderate to strong flares, as can be seen in Table \ref{table1}. Among these flares, one ($\approx$ 2.3 \%) is B-class, 12 ($\approx$ 27.3 \%) are C-class, 24 ($\approx$ 54.5 \%) are M-class, and six ($\approx$ 13.6 \%) are X-class flares. The remaining one event is associated with a strong flare (observed by the Solar TErrestrial RElations Observatory) but it is located behind the limb, when viewed from Earth. Thus, there is no GOES class for this event.
The plots of flare rise time, decay time, and flare duration versus number of events are presented in panels (a), (b), and (c) of Figure \ref{fig:flare}, respectively. The rise time, decay time, and duration of the flare are calculated as the difference between start and peak times, peak and end times, and start and end times, respectively. From the figure, we see that the majority of the flares ($\approx$ 80\%) are impulsive in nature, i.e., their rise time is within 20 minutes. Also, many events ($\approx$ 57 \%) have a decay time within 10 minutes. In case of the duration of these flares, $\approx$ 12 \% of which are within 10 minutes, $\approx$ 58 \% are within 11 -- 30 minutes and rest are in 31 -- 60 minutes time range, respectively. 

\begin{figure}[!t] 
    \centering
    \includegraphics[width=0.48\textwidth]{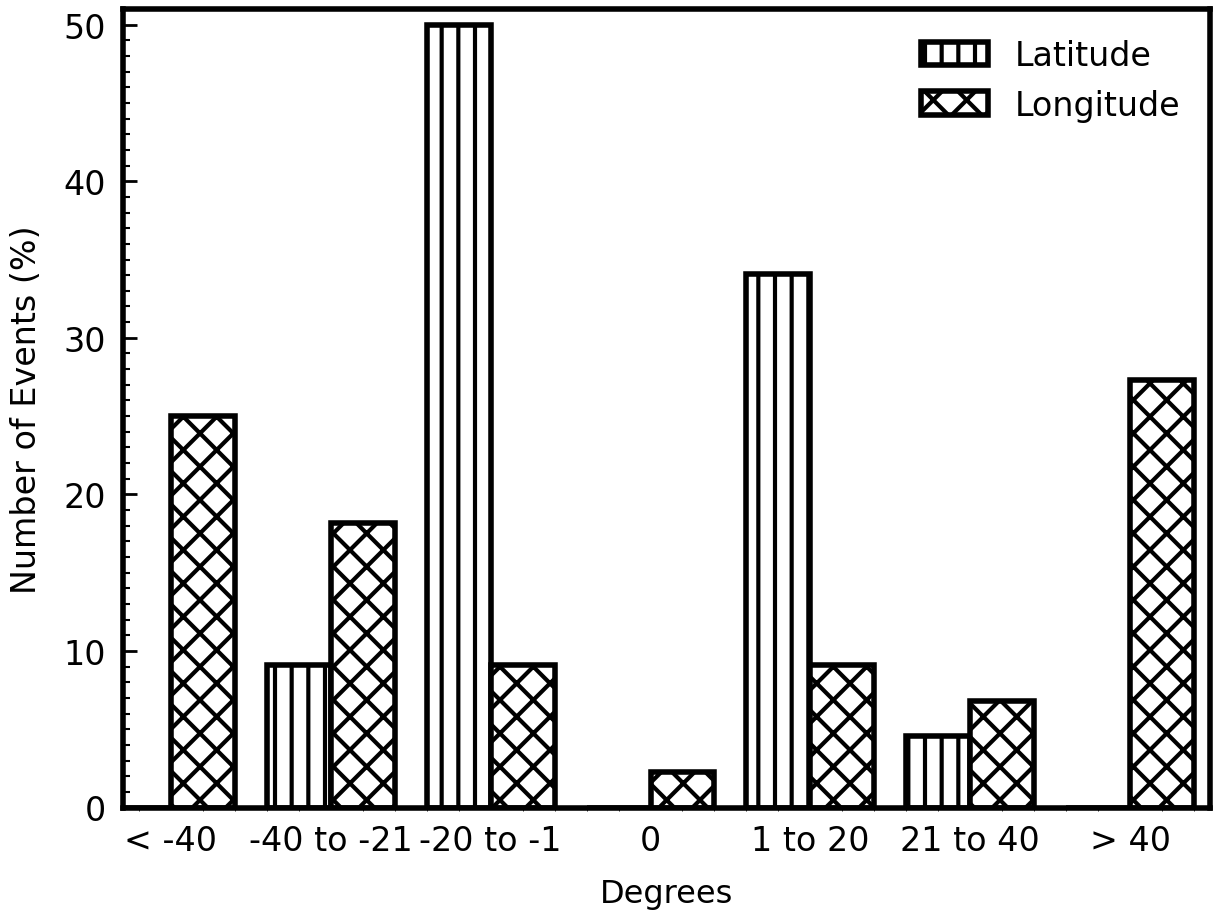}
    \caption{Distribution of location of the sources of m-type II radio bursts on the solar disk. The bars with ``vertical lines'' and ``crosses'' represent the latitude and longitude of the source location, respectively. 
    }
    \label{fig:lat_lon}
\end{figure}

\begin{figure}[!h] 
    \centering
    \includegraphics[width=0.48\textwidth]{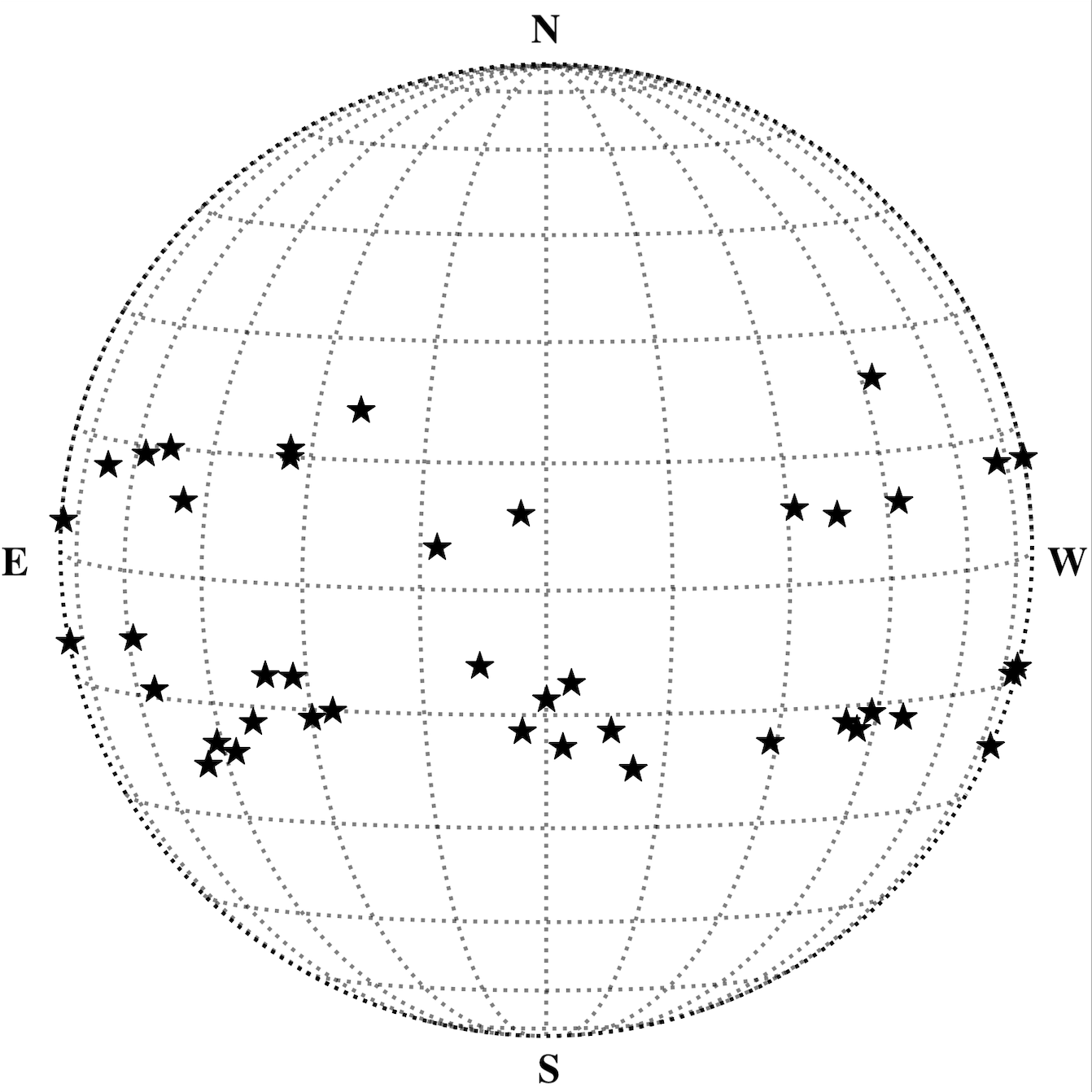}
\caption{Locations of m-type II burst sources shown on the solar disk with start symbols.}
    \label{fig:lat_lon_disk}
\end{figure}

Next, we find the source of the m-type II radio bursts on the solar disk by investigating the location of associated flares. The distribution of the source latitudes and longitudes are shown in Figure \ref{fig:lat_lon}. From the figure, we see that the type II bursts occur within the latitude of $\pm$ 30$^{\circ}$, whereas they originate from all the longitudes that cover from 0$^{\circ}$ to $\pm$ 90$^{\circ}$. Many events ($\approx$ 51 \%) lie in the bin of $-$1$^{\circ}$ to $-$20$^{\circ}$ latitude and reach maximum up to $-$24$^{\circ}$ latitude (where the negative sign is selected to denote Southern latitudes in the plots). After careful examination of the events, we find that 26 events, i.e., 59 \%, have their source located in the Southern hemisphere and the remaining 17, i.e., $\approx$ 39 \%, originate from the Northern hemisphere. In one case, we could not find the location on the disk, because of a backside event (behind the limb). For the case of helio-longitudes, 23 events ($\approx$ 52 \%) are from the Eastern and 19 ($\approx$ 43 \%) are from the western hemisphere. The remaining one event has its source at 0$^{\circ}$ longitude. The source locations on the solar disk are visualized in Figure \ref{fig:lat_lon_disk} with star symbols.

\subsection{Relationship with CMEs}
\label{sec:CME}

\begin{figure*}[!t] 
    \centering
    \includegraphics[width=\textwidth]{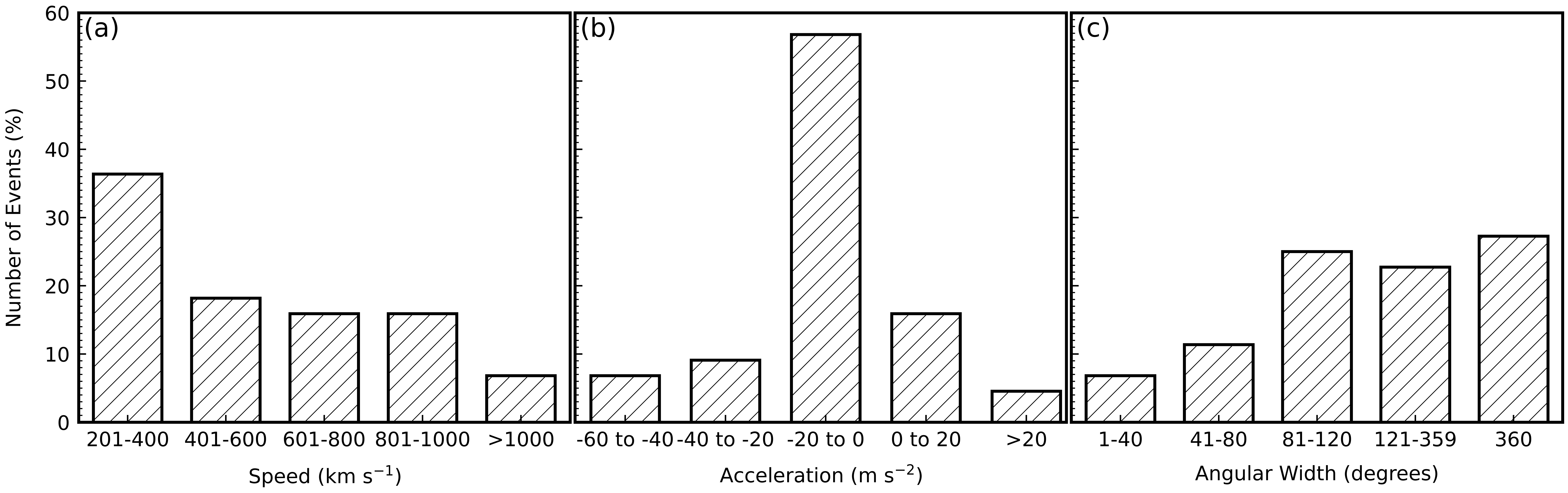}
    \caption{Histograms of speed, acceleration, and angular width of the CMEs associated with m-type II radio bursts. 
    }
    \label{fig:cme}
\end{figure*}

In order to identify the relationship of these m-type IIs with CME, we used data from the LASCO CME catalog. From there, we found that 41/44, i.e., $\approx$ 93 \% of these m-type IIs are associated with CMEs. The remaining three events are: (1) event no. 12 was originated on 25 September 2012 and was associated with a weak flare of GOES C3.6 class. (2) The event no. 20 on 25 October 2013 was associated with a failed eruption accompanied by an EUV wave, and a jet from the flaring active region. (3) The event no. 28 originated on 22 August 2014 was associated with a very weak flare of GOES B-class and a jet activity. 

From the LASCO CME catalog we take parameters such as speed, acceleration, and angular width. Their histograms are presented in Figure \ref{fig:cme}, panels (a), (b), and (c), respectively. 
From all events, $\approx$ 39 \% have CME speeds in the range 201 -- 400 \kms, and about 10 \% of CMEs have a very high speed, that is, $>$ 1000 \kms. Panel (b) of the figure shows that a large fraction of the CMEs ($\approx$ 57 \%) are decelerated up to 20 m s$^{-2}$. However, all CMEs have an acceleration ranging from $-60$ to more than 20 m s$^{-2}$. The angular width of these CMEs (Figure \ref{fig:cme}(c)) confirms that $\approx$ 27\% (12/44) of the events are halo. A similar number of events lie in the ranges of 81$^{\circ}$ -- 120$^{\circ}$ and 121$^{\circ}$ -- 359$^{\circ}$.

\subsection{Coronal and shock parameters} 
\label{sec:magnetic_field}
Several key parameters of the coronal plasma and the shock wave are calculated on the basis of our event selection. The description of these parameters are given in the following subsections.

\subsubsection{Density} 
\begin{figure}[!h]
    \centering
    \includegraphics[width=0.45\textwidth]{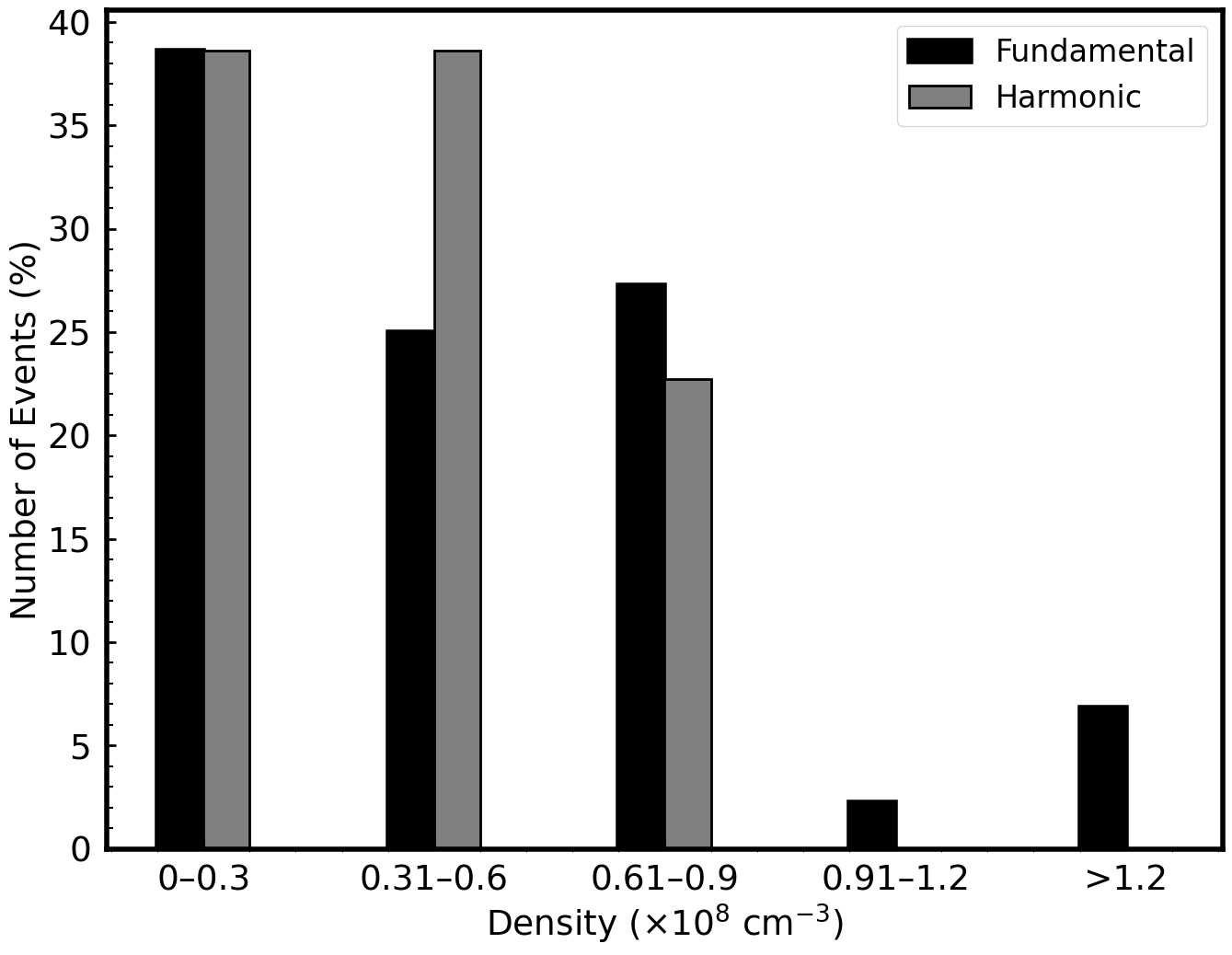}
    \caption{Histogram of density with number of events. The densities computed using the fundamental and harmonic bands are plotted with black and gray colors, respectively.}
    \label{fig:density}
\end{figure}

The density is directly related to the frequency with the following expression,
   \begin{equation}
  \label{eq:density}
  f[MHz] = 9 \times 10^{-3} \times \sqrt{n} ~;
\end{equation}
\begin{center}
$n[10^{-8}~cm^{-3}] = \Big(\frac{f}
{9\times10^{-3}}\Big)^2$
\end{center}

Figure \ref{fig:density} shows the histogram of the density calculated from the fundamental and harmonic bands using equation \ref{eq:density}. The averaged density derived from the fundamental bands is higher (up to 1.9 $\times$ 10$^8$ cm$^{-3}$) than that of the harmonic bands (maximum 0.75 $\times$ 10$^8$ cm$^{-3}$). 
In case of the density computed from fundamental bands, the distribution of events in the different density ranges is as follows: 
\begin{itemize}
    \item $\approx$ 39 \% in 0 -- 0.3 $\times$ 10$^8$ cm$^{-3}$
    \item $\approx$ 25 \% in 0.31 $\times$ 10$^8$ -- 0.6 $\times$ 10$^8$ cm$^{-3}$
    \item $\approx$ 27 \% in 0.61 $\times$ 10$^8$ -- 0.9 $\times$ 10$^8$ cm$^{-3}$
    \item $\approx$ 9 \% in 0.91 $\times$ 10$^8$ -- 2 $\times$ 10$^8$ cm$^{-3}$.
\end{itemize}
However, in case of the harmonic bands, nearly equal numbers of events can be found in two density bins, 
namely:
\begin{itemize}
    \item $\approx$ 38 \% in 0 -- 0.3 $\times$ 10$^8$ cm$^{-3}$
    \item $\approx$ 38 \% in 0.31 $\times$ 10$^8$ cm$^{-3}$ -- 0.6 $\times$ 10$^8$ cm$^{-3}$
    \item $\approx$ 24 \% in 0.61 $\times$ 10$^8$ -- 0.9 $\times$ 10$^8$ cm$^{-3}$.
\end{itemize}

\subsubsection{Type II speed and height of shock formation}
\begin{figure*} 
    \centering
    \includegraphics[width=0.48\textwidth]{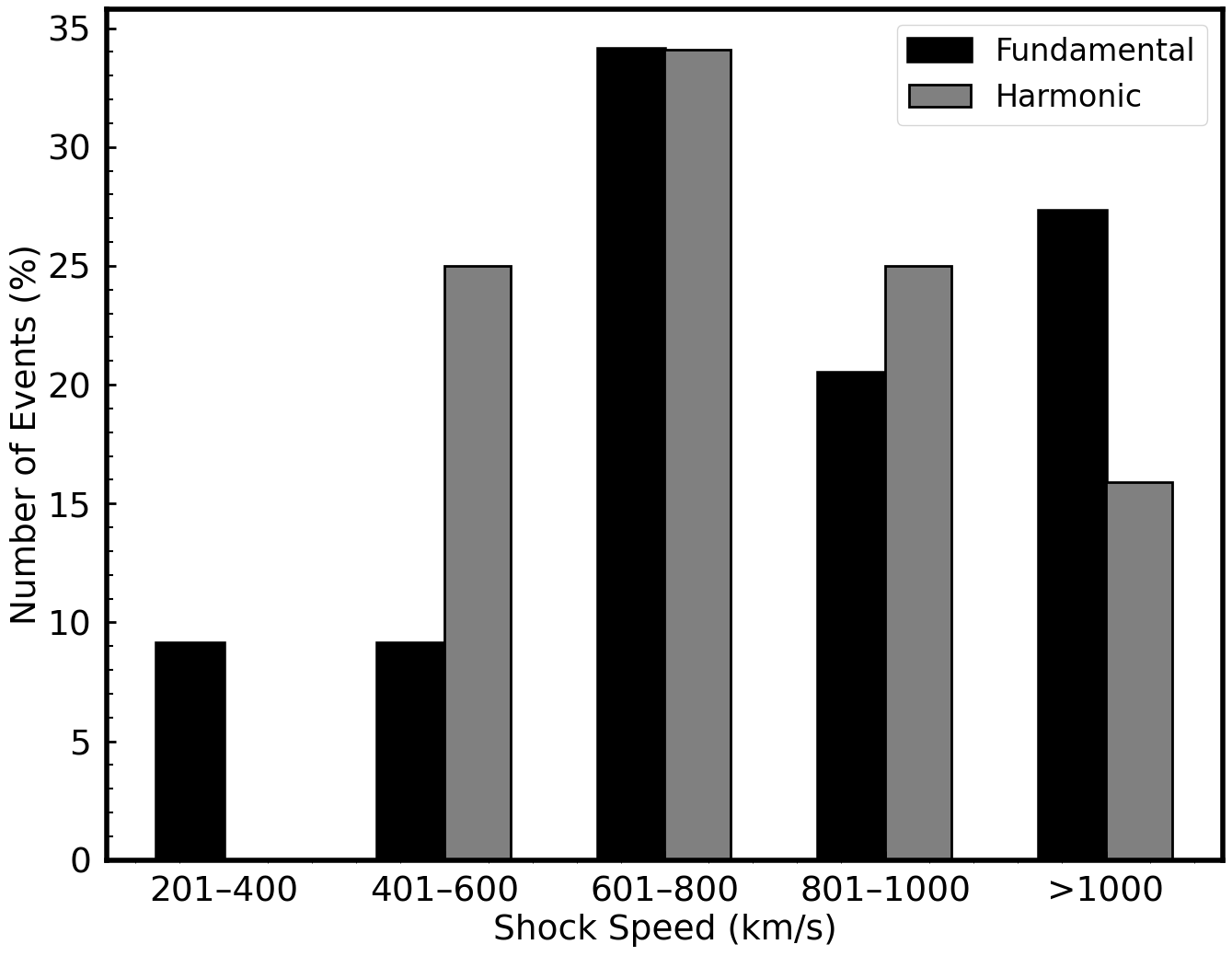}
    \includegraphics[width=0.48\textwidth]{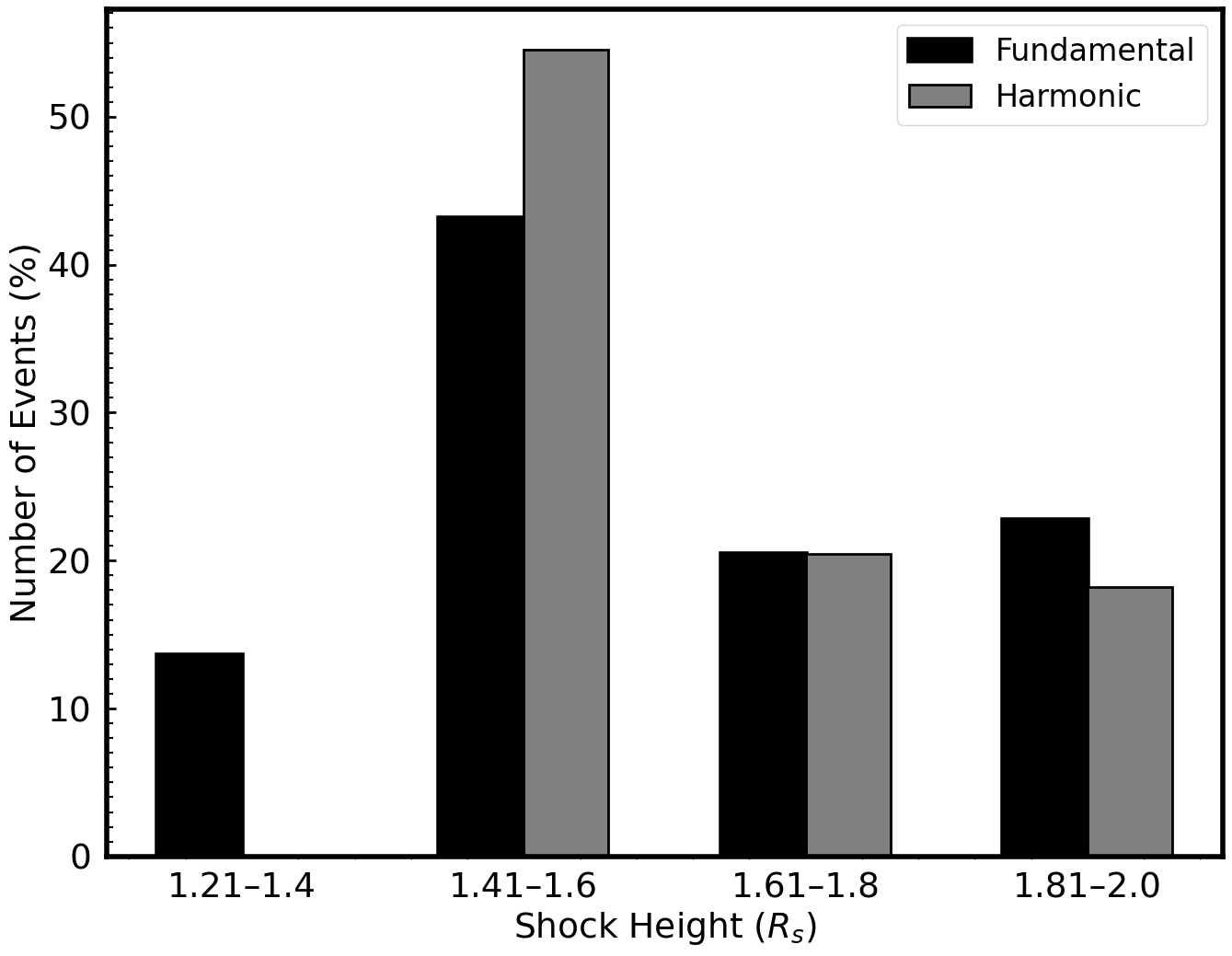}
    \caption{Distribution of the speed (left) and height (right) of m-type II radio bursts computed from the hybrid model. The speed and height using fundamental and harmonic bands are depicted with black and gray bars, respectively. }
    \label{fig:shock_speed}
\end{figure*}

The speed of type II radio bursts is determined here through the application of the  ``hybrid'' density model, which was introduced by \citet{Vrsnak2004}. In their study, the authors evaluated the performance of pre-existing electron density models, specifically the two-fold Saito model \citep{Saito1970} and the Leblanc model \citep{Leblanc1998}, in terms of their ability to represent the full range of type II radio bursts from the coronal to IP regions. Their findings indicate that neither of these conventional models, whether used individually or in combination, were adequate for accurately characterizing the density in both coronal and IP domain.

To address this limitation and provide a more consistent and realistic representation of the electron density throughout the entire region of interest, \citet{Vrsnak2004} developed the hybrid model. This model was constructed to offer a smooth and continuous transition in the density profile, starting from the low-corona regions associated with active regions and extending outward into the IP medium. One of the key features of the hybrid model is its compatibility with the magnetic field scaling relationship, $B\propto R^{-2}$ (where $R$ is the radial distance), which becomes particularly valid at heliocentric distances beyond $R \approx 2 R_s$. This feature ensures that the model remains consistent with observed solar wind magnetic field behavior in the outer corona and IP space.
The density model better reflects the physical conditions encountered by type II radio bursts as they propagate through different regions of the solar atmosphere. The final formulation of this newly constructed hybrid model is given by \citep{Vrsnak2004}:

\begin{equation}
\label{eq:hybrid}
    n[10^8 ~cm^{-3}] = \frac{15.45}{R^{16}} + \frac{3.16}{R^6} + \frac{1}{R^4} + \frac{0.0033}{R^2}
\end{equation}

After computing the $R$ using density ($n$) from equations \ref{eq:density} and \ref{eq:hybrid}, we found the speed of the type II radio bursts as

\begin{equation}
   v = \frac{dR}{dt} = \frac{R_{i+1} - R_{i}}{t_{i+1}-t_{i}}  
\end{equation}
where $t$ is the time.

Figure \ref{fig:shock_speed} presents the distribution of shock speed and the shock formation height. The figure is divided into two panels for clarity and comparative analysis. The left panel illustrates the distribution of shock speeds derived from the observed m-type II radio bursts, while the right panel shows the corresponding distribution of shock heights at the onset of these bursts.

The shock speeds (left panel) have been estimated separately for m-type II bursts identified in both the fundamental (black bars) and harmonic (gray bars) bands. The calculated shock speeds span a wide range, from approximately 359 \kms\ to 1687 \kms\ for the fundamental, and from about 408 \kms\ to 1727 \kms\ for the harmonic counterparts. It is noticed that nearly one third of the events, i.e., $\approx$ 34 \% in both fundamental and harmonic categories fall within the 601 -- 800 \kms\ range, indicating that this is the most frequently occurring shock speed range among the analyzed m-type II bursts.

The heights of shock formation (right panel of Figure \ref{fig:shock_speed}) are determined from the frequency at the LFB of the type II burst at the time of start of type II burst. For the fundamental frequency band, the calculated shock heights range from 1.21 R$_s$ to 2.0 R$_s$, while for the harmonic band, the corresponding range is slightly narrower, extending from 1.41 R$_s$ to 2.0 R$_s$. The analysis shows that the majority of the events are concentrated in the 1.41 -- 1.6 R$_s$ range. Specifically, $\approx$ 43 \% of the fundamental and an even higher fraction, 55 \%, of the harmonic band events occur within this height interval. This suggests that a significant number of type II bursts are generated at relatively low coronal heights, just above the solar surface, particularly around 1.5 R$_s$.

\begin{figure} 
    \centering
    \includegraphics[width=0.47\textwidth]{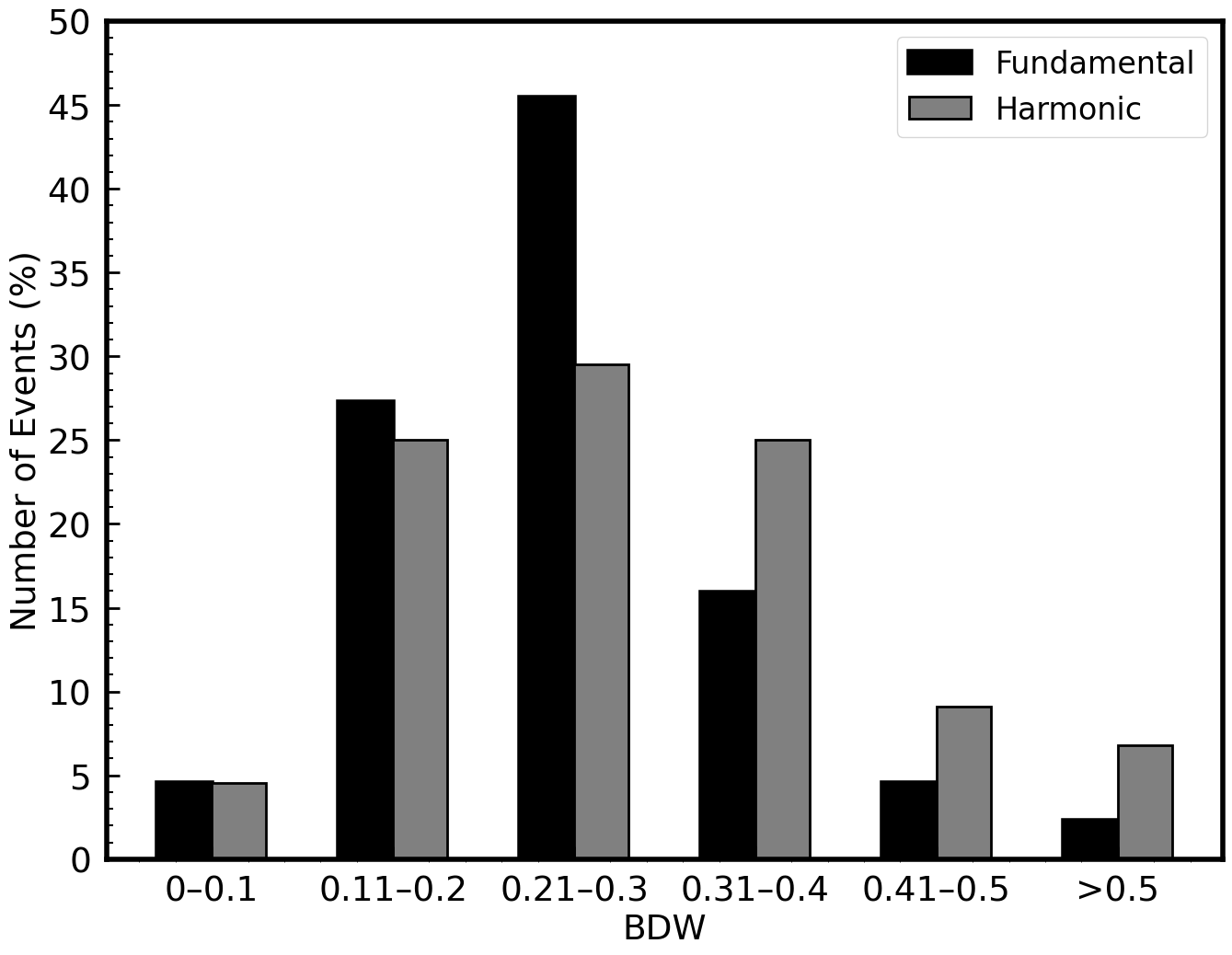}
    \caption{Distribution of BDW computed from the fundamental (black bars) and harmonic (gray bars) bands. }
    \label{fig:bdw}
\end{figure}

\subsubsection{Relative band-splitting (BDW)}
In our current study, we are using the m-type II radio bursts in which the fundamental and harmonic bands shows splitting. The parameters of this splitting, such as the frequencies and times of the UFBs and LFBs, are used to calculate other coronal parameters. Their values are summarized in Table~\ref{table2}. The expressions used and the computed parameters are given below.

First, we calculate the relative band-splitting (BDW) which is given by,
\begin{equation}
\label{eq:bdw}
    BDW = \frac{\Delta f}{f} = \frac{(f_U-f_L)}{f_L}
\end{equation}

The distribution of relative BDW is presented in Figure \ref{fig:bdw} with black and gray bars for the fundamental and harmonic bands, respectively. In case of BDW computed from fundamental bands, we obtain the following distribution of events in the respective BDW bins:
\begin{itemize}
    \item $\approx$ 45 \% in 0.21 -- 0.3
    \item $\approx$ 27 \% in 0.11 -- 0.2
    \item $\approx$ 16 \% in 0.31 -- 0.4
    \item $\approx$ 12 \% in 0 -- 0.1 and 0.41 to $>$ 0.5
\end{itemize}
Similarly, the majority of the BDW values computed from harmonic bands lie in 0.11 -- 0.4, with the following detailed distribution:
\begin{itemize}
    \item $\approx$ 25 \% in 0.11 -- 0.2
    \item $\approx$ 30 \% in 0.21 -- 0.3
    \item $\approx$ 25 \% in 0.31 -- 0.4 
    \item $\approx$ 20 \% in 0.41 to $>$ 0.5
\end{itemize}
The average value of BDW are 0.25 and 0.29, respectively, for fundamental and harmonic bands.

\begin{figure} 
    \centering
    \includegraphics[width=0.48\textwidth]{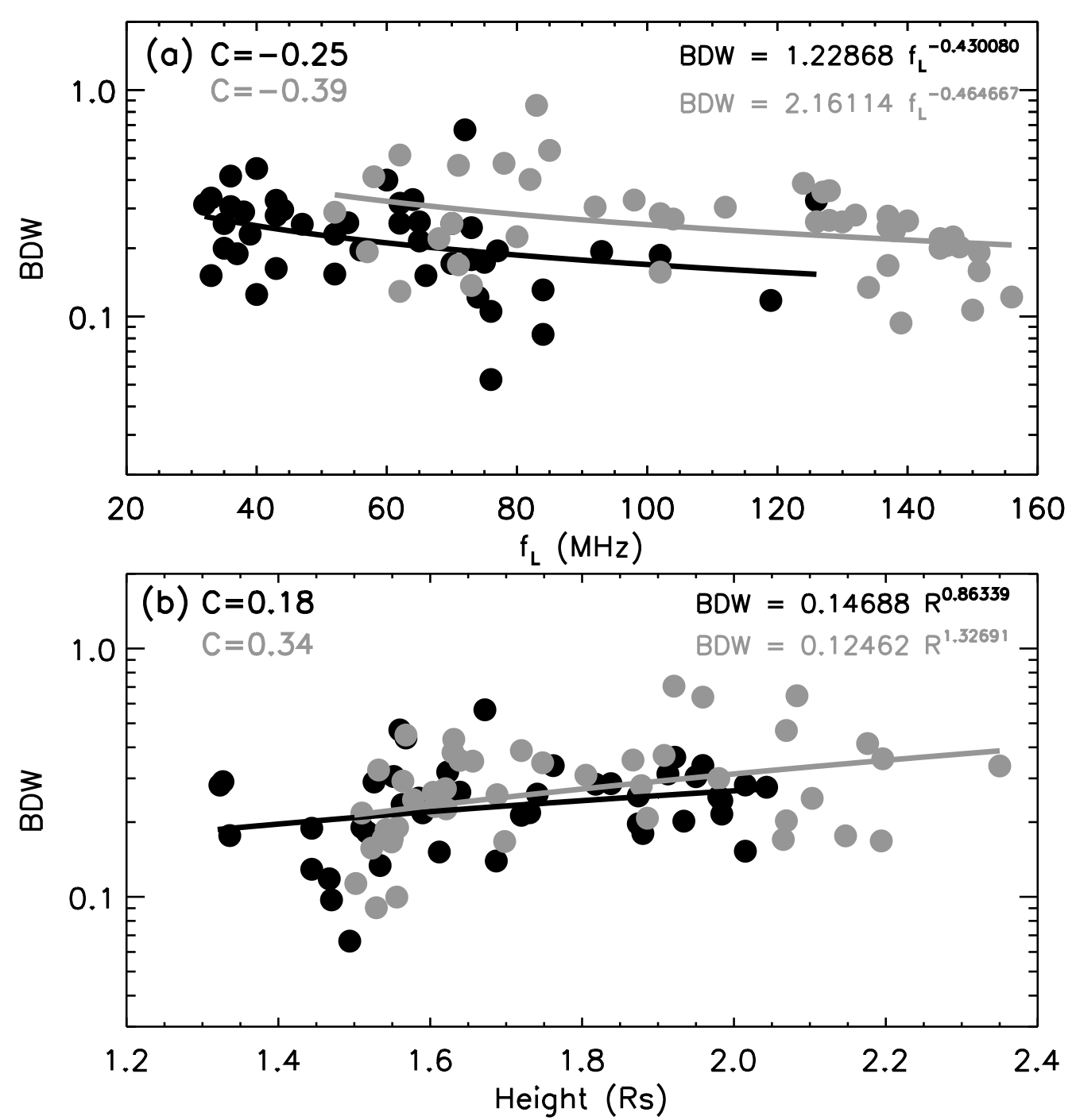}
    \caption{Scattered plots of BDW versus LFB frequency (top) and radial height of solar corona (bottom). The black and gray dots denotes the BDW values computed from fundamental and harmonic bands, respectively. The black and gray solid lines represents the power low fitting in the data points from fundamental and harmonic bands, respectively. The values of `C' at the top left of each panel is the Pearson correlation coefficients and at top right is the fitted power law.}
    \label{fig:bdw_fl_rad}
\end{figure}

Figure \ref{fig:bdw_fl_rad} displays the scattered plots of BDW with LFB frequency, i.e., $f_L$ and height of solar corona in panels (a) and (b), respectively. The black and gray colors are chosen for the BDW values from the fundamental and harmonic bands, respectively. To check the relationship of BDW with $f_L$ and the height of the solar corona, we compute the correlation of BDW with these quantities. We find that BDW is very weakly correlated with $f_L$ with a correlation coefficient of $-0.25$ and $-0.39$ for the fundamental and harmonic bands, respectively. We also fit a power law in the scattered plots which are written on the top right of the figure. From power law fitting, we find that the BDW $\propto~f_L^{-0.430080}$ for fundamental and BDW $\propto~f_L^{-0.464667}$ for harmonic band.
On the other hand, BDW is weakly correlated to the height of the solar corona but with a positive correlation coefficient of 0.18 and 0.34 for fundamental and harmonic bands, respectively. The power law fitting in panel (b) of the figure gives BDW $\propto~R^{0.86339}$ for the fundamental and BDW $\propto~R^{1.32691}$ for the harmonic band. The low correlation coefficient suggests that the BDW is almost constant over all frequency spectrum and with the height of the solar corona.

\subsubsection{Drift rate}

\begin{figure} 
    \centering
    \includegraphics[width=0.48\textwidth]{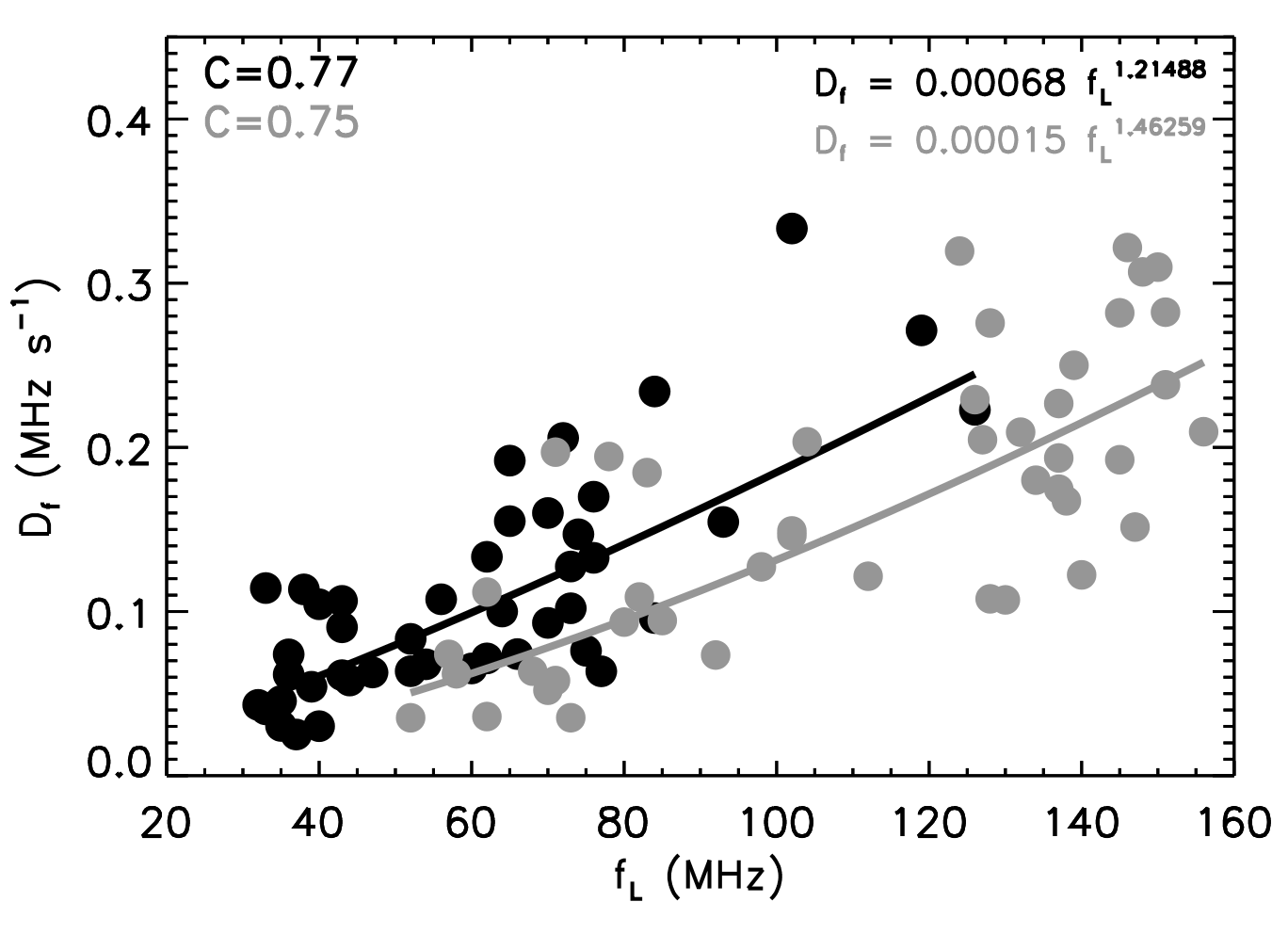}
    \caption{Variation on drift rate ($D_f$) with the LFB frequency, $f_L$ shown with black and gray dots. The solid lines are the power law fit to the data. The black and gray colors represents the same as Figure \ref{fig:bdw_fl_rad}. The `C' at the top right represent the Pearson correlation coefficients for fundamental and harmonic bands with black and gray colors, respectively.. }
    \label{fig:drift_rate}
\end{figure}

The drift rate, denoted as $D_f$, quantifies the rate of change of frequency over time and is given as:
 \begin{center}
     $D_f ~ = ~ \frac{f_{start}-f_{end}}{t_{start}-t_{end}}$
 \end{center}

where $f_{start}~ (f_{end})$ and $t_{start}~ (t_{end})$ are the frequency and time at the start (end) of type II radio burst spectrum, respectively. This calculation is carried out separately for the fundamental and harmonic frequency bands, following a methodology similar to that used in the BDW analysis. After computing the values of $D_f$ for both bands, these values are plotted against the corresponding values of $f_L$, in order to examine the dependence of the drift rate on the LFB frequency. The resulting scatter plot is presented in Figure \ref{fig:drift_rate}. For consistency, the color notation in this figure is identical to that used in Figure \ref{fig:bdw_fl_rad}. From the plot, we see that $D_f$ increases with an increase in $f_L$. This trend holds for both the fundamental and harmonic frequency bands, suggesting a positive correlation between these two quantities. To further quantify this relationship, we calculated the linear (Pearson) correlation coefficients which are found to be 0.77 for the fundamental band and 0.75 for the harmonic band, indicating a strong and statistically significant correlation in both cases.

To characterize the nature of this dependency more precisely, we apply a power-law fit to the scatter plot data. The fitted relations give the dependence of $D_f$ on $f_L$ as $D_f\propto f_L^{1.21488}$ and $D_f\propto f_L^{1.46259}$ in case of fundamental and harmonic bands, respectively. These fitted equations, which provide a mathematical description of how the drift rate varies with frequency, are displayed in the upper right corner of Figure \ref{fig:drift_rate}. 

\subsubsection{Alfv\'en Mach number}

\begin{figure}[!t]  
    \centering
    \includegraphics[width=0.48\textwidth]{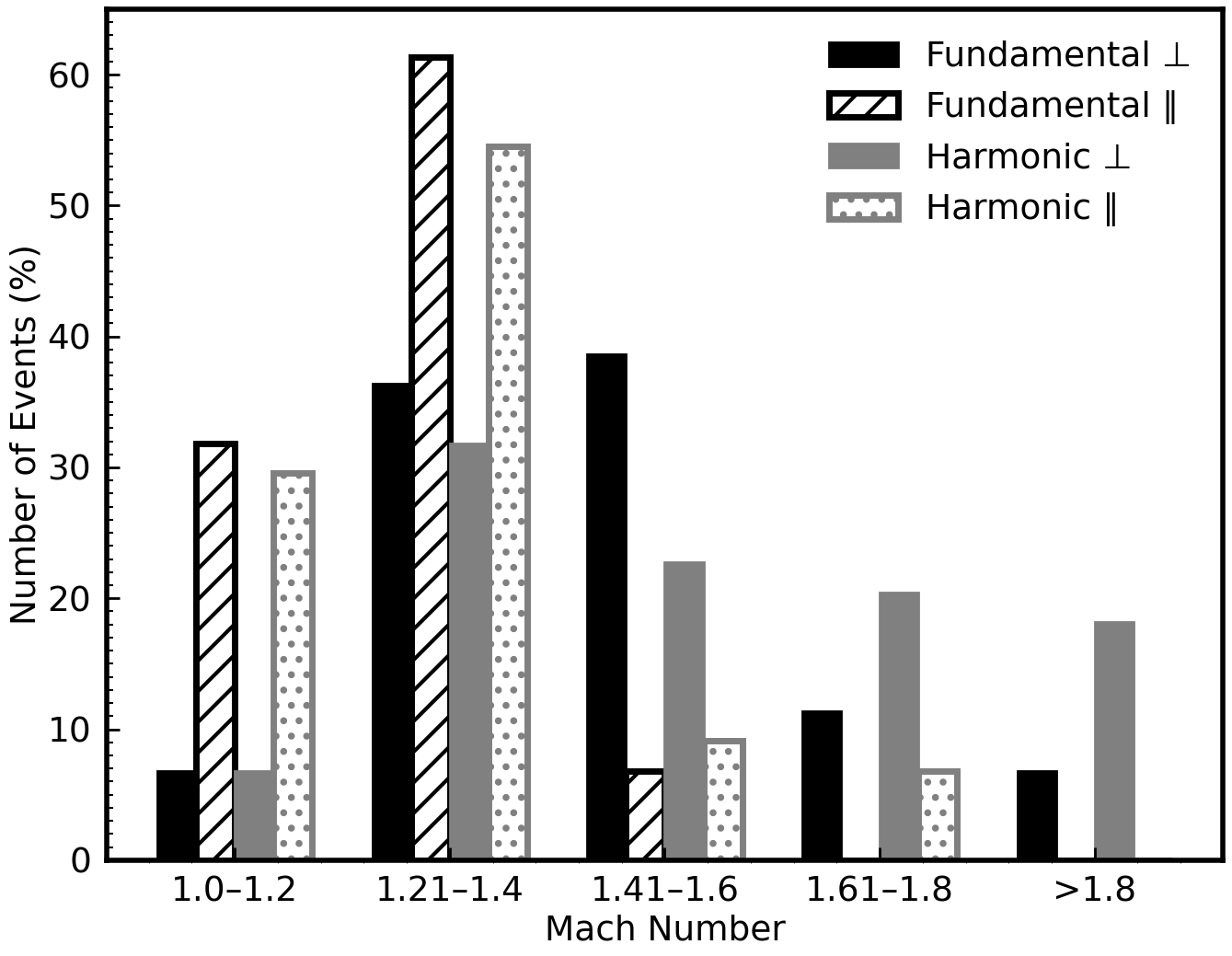}
    \caption{Histogram of mach number computed separately for fundamental and harmonic bands of m-type II radio bursts.}
    \label{fig:mach_number}
\end{figure}
The Alfv\'en Mach number ($M_A$) is computed using the expression (called Rankine-Hugoniot jump relation) suggested by \cite{Vrsnak2002}, which is given as follows:

\begin{equation}
      M_A = \begin{cases}
    \sqrt{\frac{X(X+5)}{2(4-X)}}, ~~for ~perpendicular ~shock \\
      \sqrt{X}, ~~~~~~~~~~~for ~parallel ~shock
   \end{cases}
\end{equation}

Here, X is the density jump that can be calculated using the band splitting in type II bands. It is given by the relation X = $\Big(\frac{f_U}{f_L}\Big)^2$. Now, as frequency and density are related to each other through the relation $f \propto \sqrt{n}$, presented in Equation \ref{eq:density}, using the latter and Equation \ref{eq:bdw}, the density jump (X) can be expressed as
\begin{equation}
    X = \Big(\frac{f_U}{f_L}\Big)^2 = \frac{n_2}{n_1} = (BDW+1)^2 
\end{equation} 

Using the above equations, we computed the density jump (X) and the Alfv\'en Mach number (M$_A$). This gives different values for fundamental and harmonic bands as the frequencies are different in these bands. No information is available whether the shock driver of the type II radio bursts is perpendicular or parallel. Therefore, for completeness, we computed the M$_A$ using both formulae. In this way, we obtain two values of M$_A$ for fundamental and two for harmonic bands, viz., one for perpendicular and one for parallel shock, i.e., in total four values of M$_A$ for a single event. The histogram of the computed M$_A$ with the number of events is presented in Figure \ref{fig:mach_number}. The four different values are plotted in different color-coded bars, as explained in the figure legend, namely, the black solid bar and the bar with black slant lines represent the M$_A$ derived from the fundamental band considering perpendicular and parallel shocks, respectively. Similarly, solid gray bar and the bar with gray dots represent the M$_A$ derived from the harmonic band considering the perpendicular and parallel shocks, respectively. 

Below we explain the distribution of fundamental and harmonic bands with perpendicular and parallel shocks. In the case of M$_A$ calculated from the fundamental band with perpendicular shock (black solid bars), the majority of the events ($\approx$ 82 \%) lie in the range 1 -- 1.6. The remaining events are in the range 1.61 -- 2.5, with the maximum value of M$_A$ $\approx$ 2.5. For parallel shocks (bars with black slant lines), $\approx$ 93 \% of the events have M$_A$ in the range 1 -- 1.4. The maximum value of M$_A$ in this case is $\approx$ 1.57. 

On the other hand, the M$_A$ computed from the harmonic band with perpendicular shock (gray solid bars) is distributed as follows: $\approx$ 32 \% for this case lie in the range 1.21 -- 1.4 and almost an equal number of events are distributed in the range 1.41 -- 1.6 ($\approx$ 20 \%) and 1.61 -- 1.8 ($\approx$ 22 \%).  
A similar distribution is shown by the M$_A$ calculated from parallel shock (gray bars with dots), in the sense that most of events ($\approx$ 55 \%) lie in the range of 1.21 -- 1.4. However, the number of events for the smallest values of M$_A$, 1.0 -- 1.2, is significantly higher, about 30 \%, which is $\approx$ 1.7 times more compared to the case of perpendicular shock. The maximum value of M$_A$ in this case is 1.7.

\subsubsection{Alfv\'en speed}
\begin{figure}[!t] 
    \centering
    \includegraphics[width=0.45\textwidth]{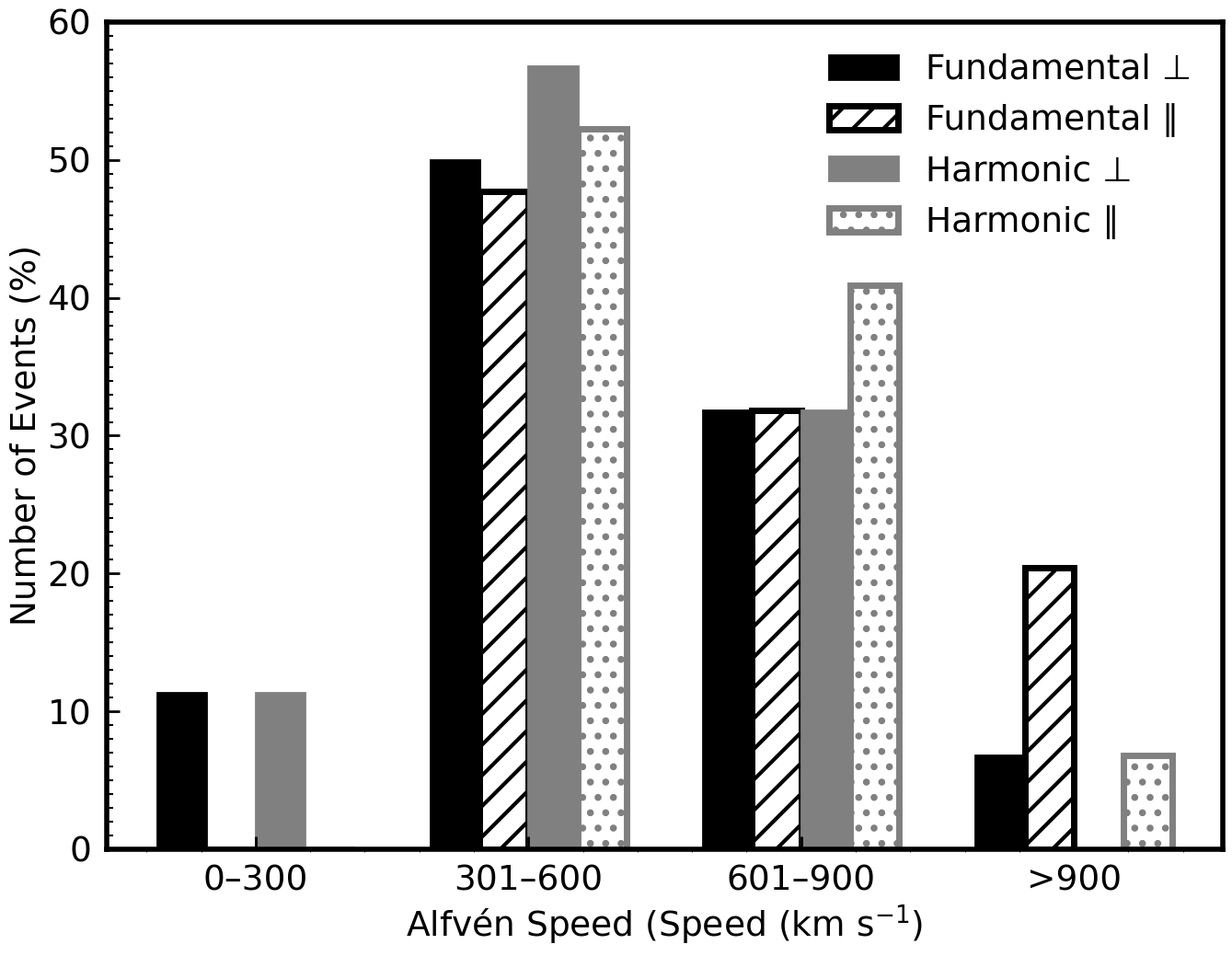}
    \caption{Histograms of Alfv\'en speed computed from fundamental and harmonic bands of m-type II radio bursts.}
    \label{fig:alfven_speed}
\end{figure}

After the computation of M$_A$, its values are used to calculate the Alfv\'en speed using the following equation:
\begin{equation}
\label{eq:alfven}
    v_A = \frac{v}{M_A}
\end{equation}
where $v$ is the speed of the type II radio bursts, i.e., shock speed. We calculated the Alfv\'en speed separately for the fundamental and harmonic bands, considering both types of shocks, viz., perpendicular and parallel, shown in Figure \ref{fig:alfven_speed}. The color-code used, namely, solid black and gray bars, black bars with slant lines, gray bars with dots, is similar to that in Figure \ref{fig:mach_number}. 

From the figure, we see that the majority of events in each case, i.e., $>$ 48 \%, have the Alfv\'en speed in the range 301 -- 600 \kms. The bin of 601 -- 900 \kms\ have more than $\approx$ 32 \% of events in this range. As far as the minimum and maximum values of Alfv\'en speed is concerned, it ranges from 267 -- 1071 \kms for fundamental bands with perpendicular shock, 301 -- 1294 \kms for fundamental bands with parallel shock, 233 -- 861 \kms for harmonic bands with perpendicular shock, and 316 -- 1177 \kms for harmonic bands with parallel shock.

\subsubsection{Coronal magnetic field}
\begin{figure}[!t] 
    \centering
    \includegraphics[width=0.45\textwidth]{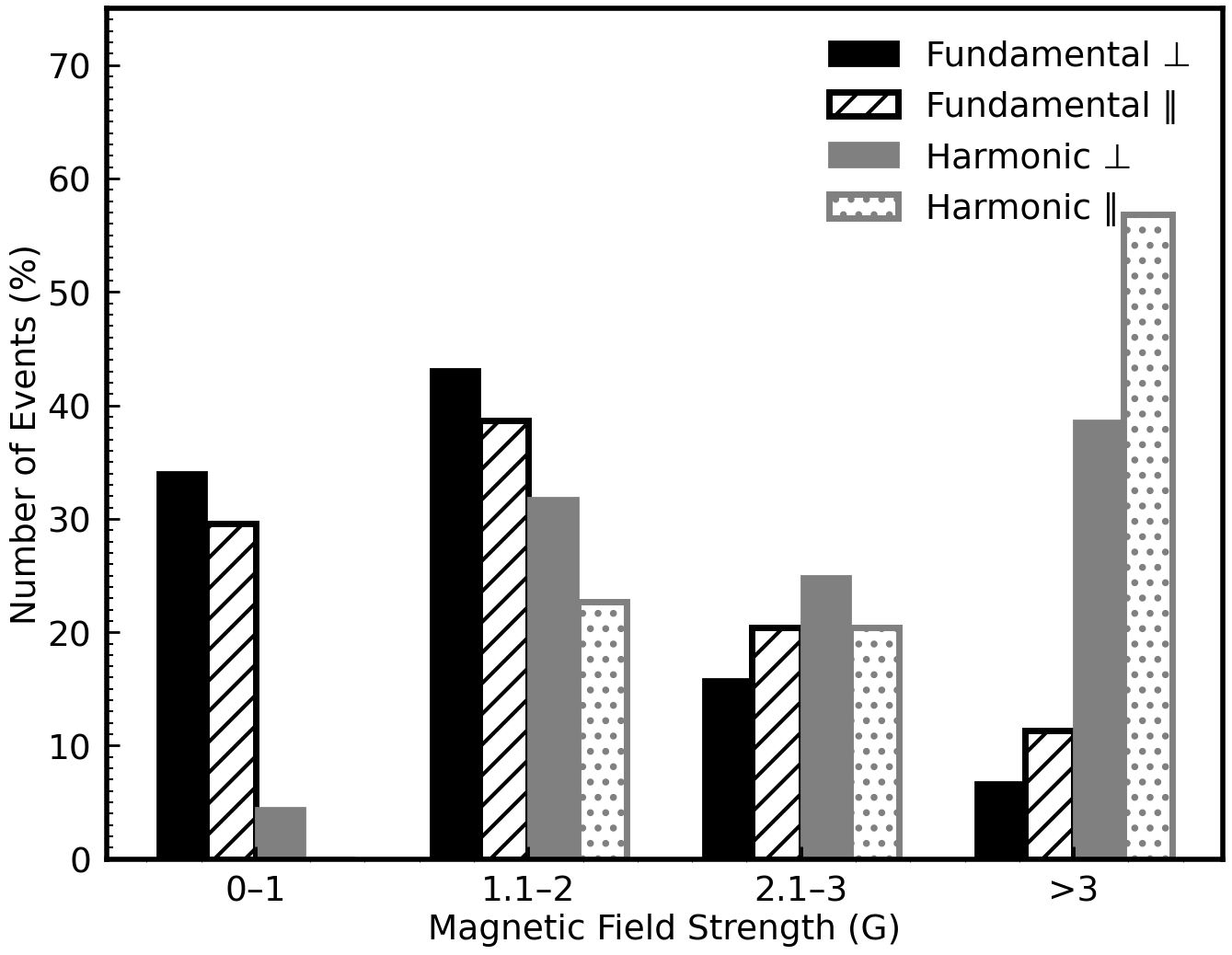}
    \caption{Distribution of magnetic field strength calculated for perpendicular and parallel shock using fundamental and harmonic bands in type II bursts. 
    }
    \label{fig:magnetic_field}
\end{figure}

Insofar, the values of all the coronal parameters calculated in the above subsections are present. These parameters are used to find the coronal magnetic field strength using the following expression \citep{Vrsnak2002, Gopalswamy2012case, Vasanth2014},
\begin{equation}
\label{eq:magnetic_field}
    B = 5.1 \times 10^{-5} \times f \times v_A
\end{equation}
where the magnetic field $B$ is in Gauss (G), $f$ and $v_A$ in MHz and \kms, respectively. The frequency is taken as the average of the frequencies at the beginning and end of the type II burst, that is, $f = (f_{start}+f_{end})/2$. 
As we have the values of $v_A$ from the fundamental band perpendicular and parallel shocks, and the harmonic band perpendicular and parallel shocks, we get four values of $B$ for a single event. 

The distribution of the computed values of $B$ using equation \ref{eq:magnetic_field} are plotted in Figure \ref{fig:magnetic_field}. The colors of the bars in left panel have the same meaning as in Figures \ref{fig:mach_number} and \ref{fig:alfven_speed}. The $B$ computed from the fundamental bands is lower than that from the harmonic bands. From the figure, it is seen that the majority of events ($>$ 40 \%) have $B$ in range of 1.1 -- 2 G, both in case of perpendicular and parallel shock. However, in the case of harmonic bands, $B$ is $>$ 3 G for the majority events ($>$ 39 \%). 

\begin{figure*} 
    \centering
    \includegraphics[width=0.9\textwidth]{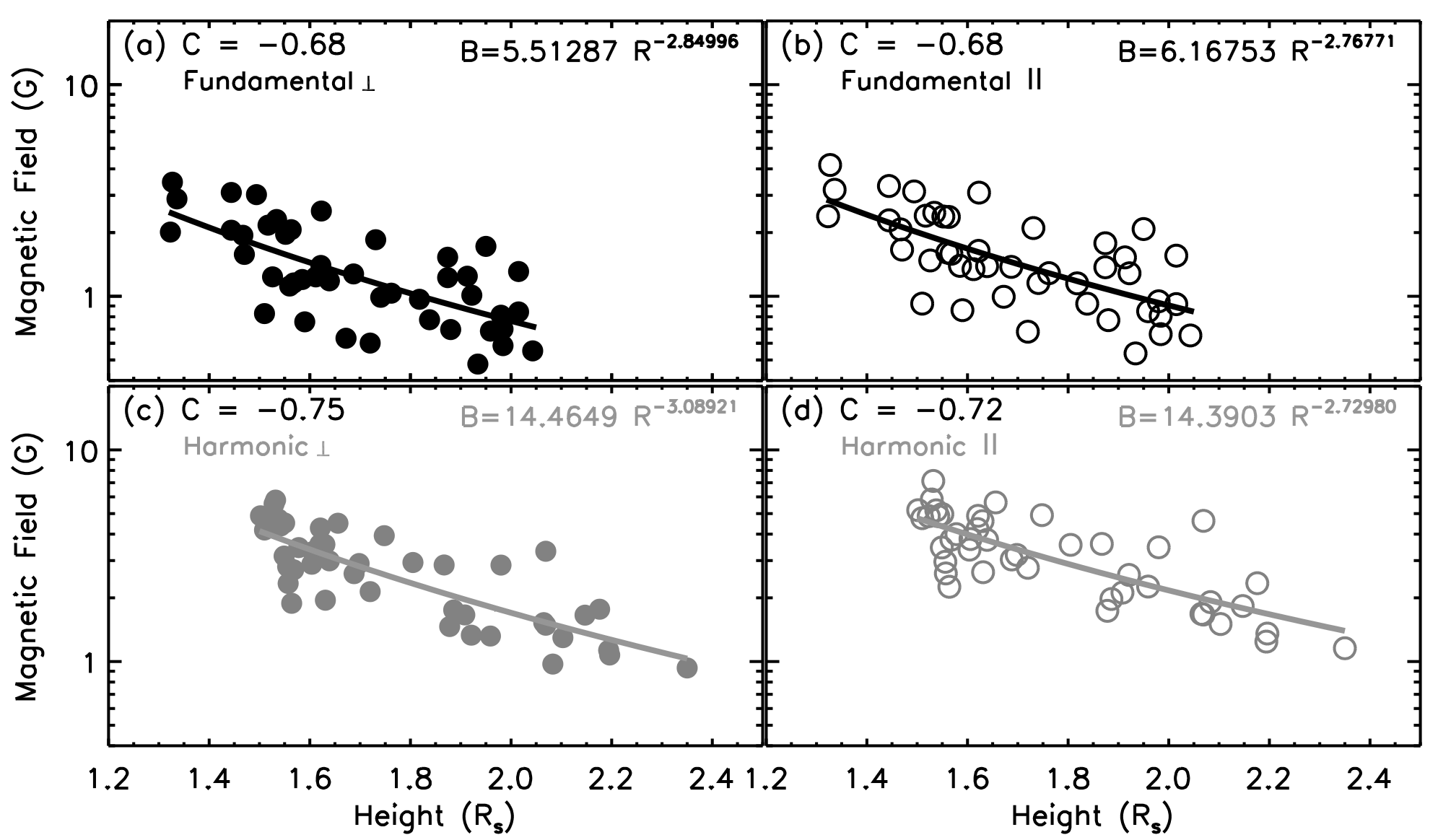}
    \caption{Variation of coronal magnetic field strength with radial distance from the Sun computed from fundamental and harmonic bands in top and bottom panels, respectively. The filled and empty circles represent the magnetic field strength from the perpendicular and parallel shocks, respectively. The power law fit are shown with solid lines in each panel.}
    \label{fig:magnetic_field_rad}
\end{figure*}

We also plotted the variation of $B$ with the radial distance of the Sun in Figure \ref{fig:magnetic_field_rad}. The filled dots (panel (a)) and empty circles (panel (b)) denote the $B$ computed from the fundamental band with perpendicular and parallel shock, and gray dots (panel (c)) and gray empty circles (panel (d)) from the harmonic with perpendicular and parallel shock, respectively. This figure again confirms that the value of $B$ from fundamental bands is less than that from the harmonic bands. 

Overall, $B$ decreases with the height of solar corona, i.e., with an increase in radial distance from the Sun, regardless of fundamental or harmonic band, and perpendicular or parallel shock. In order to quantify the correlation between the computed magnetic field and the radial distance, we used the linear (Pearson) correlations. We found a negative correlation coefficient of $\approx$ $-0.68$, $-0.68$, $-0.75$, and $-0.72$ in case of magnetic field calculated from fundamental using a perpendicular shock, fundamental using parallel shock, harmonic using a perpendicular shock, and harmonic using parallel shock, respectively. In summary, the magnetic field is inversely correlated to the radial distance of the Sun. Finally, the power law fitting to these data led to $B \propto R^{-2.85}$, $B \propto R^{-2.77}$, $B \propto R^{-3.09}$, and $B \propto R^{-2.73}$, respectively.

\section{Discussion}
\label{sec:discussion}

\subsection{CMEs and flares as type II burst drivers}
It is believed that when a CME moves fast enough, it can drive a shock wave ahead of it. This occurs when the CME surpasses the local Alfv\'en speed -- the characteristic speed at which magnetic disturbances travel through the solar plasma. Previous studies have reported a robust association between DH-type II bursts and CMEs at 1 AU \citep{Sheeley1985}. However, in the lower corona, the connection between m-type II bursts and CMEs appears less consistent \citep{Sheeley1985, Gopal1998, Reiner1999, Gopal2008}.
Our present analysis specifically investigates m-type II radio bursts that exhibit a band-splitting in both the fundamental and harmonic components. Band-splitting is often considered a signature of strong shocks. Our results indicate that 41 out of 44 m-type II events ($\approx$ 93 \%) are associated with CMEs. This high association rate implies that strong m-type II bursts -- particularly those with band-splitting -- are closely linked to CME-driven shocks.

Furthermore, \citet{Michalek2007} proposed that fast and wide CMEs are more likely to be associated with type II radio bursts compared to their slower, narrower counterparts. Our findings support this interpretation. The substantial majority of the CMEs in our sample are wide CMEs, among them 12 are halo and 10 are partial-halo \citep[i.e., angular width $>$ 120$^{\circ}$ but $<$ 360$^{\circ}$, as defined in][]{Gopalswamy2010halo} CMEs, as illustrated in Figure~\ref{fig:cme}(c). 
These halo events exhibit high velocities, ranging from 555 to 2625 \kms, whereas non-halo CMEs tend to fall within a slower speed range of 215 to 830 \kms. This supports the conclusion that fast, wide (halo) CMEs are more effective at generating strong shocks capable of producing type II radio emissions. 

All of m-type II cases in our study were also associated with solar flares, thus a flare-contribution cannot be excluded in either case. As far as the location of type II sources on the solar disk is concerned, it is found that majority of events are originated from the southern hemisphere (see Figure \ref{fig:lat_lon_disk}), which confirms the north-south asymmetry \citep{Carbonell1993}. This is due to the southern dominance of solar activity in solar cycle 24, as reported in previous studies using different data sets, namely, flares, CMEs, Sunspots, solar energetic particles, etc. \citep{Chandra2013, Li2019, 2021MNRAS.505.5212S, Zhang2023}.

\subsection{Non-CME drivers of type II bursts}
To explain the remaining two m-type II radio bursts without CMEs, there can be other possibilities drives. These possibilities include EUV waves, and small scale jet events \citep{Magdalenic2008, Pohjolainen2008, Magdalenic2012, Maguire2021, Morosan2023}. In our case, we have found three events (event no. 12, 20, and 28 in Table \ref{table1}) that are not associated with CMEs. In event no. 12, the type II radio burst could be produced by the associated flare only, i.e. the type II bursts is clearly due to a flare pressure pulse, since a CME could not be isolated.
\cite{Morosan2023} did a detailed study of event no. 20 and found that this type II radio burst is generated by coronal wave, that steepens into a shock wave, associated with the C-class flare. The event no. 28 was associated with a very weak flare of B-class. Therefore, there is a high chance that this type II bursts is not produced by the associated flare. On careful examination of the event, we find that a jet is associated with this flare. Therefore, this jet activity can be a possible generation of the type II shock.\\

\subsection{Type II band splitting and BDW}

For the computation of BDW, \cite{2020AstL...46..144T} took the pixel coordinates and then converted them into digital values. In our case, the estimation of the type II properties is done by eye and is based on the visible length of the type II bands, namely by taking values at the start and end of the spectra. In such a way, the results are an approximation and could be considered as averaged values, representative for the entire duration of the bursts.

The values of BDW in our study range from 0.06 -- 0.57 and 0.09 -- 0.70 in the case of fundamental and harmonic bands, respectively. The average value of BDW for fundamental bands is 0.25 and for harmonic band it is 0.29. In \cite{Mann1996}, the average value of BDW is 0.32 in m-domain. Moreover, in DH-km domain studied by \cite{Aguilar2005} the value of BDW for type II radio bursts from different instruments varies from 0.26 -- 0.5. Our values of BDW are slightly different from that of \cite{Mann1996} but are well in range of \cite{Aguilar2005}. Also, from Figure \ref{fig:bdw_fl_rad}, we see that the BDW does not change much with the frequency (or wavelength) and radial distance. This is because the change in frequency depends on the change in density as $\Delta f/f = \Delta n/2n$ \citep{Lengyel1989}, and the density fluctuation structure is almost the same for the entire spectral domain \citep{Vrsnak2001, Vrsnak2002, Aguilar2005}. 

The power law fitting in the scattered plots of BDW with frequency ($f_L$) and radial distance (R) suggest that the BDW varies as $BDW \propto~ f_L^{\approx -0.4}$ and $BDW \propto~R^{\approx 1}$. This is consistent to the results obtained in previous studies \citep[for example,][]{Vrsnak2001, Vrsnak2004, Aguilar2005}.

\subsection{Drift rate}

The frequency drift rate ($D_f$) of type II radio bursts characterizes how the emission frequency changes with time, reflecting the motion of the shock through the solar corona or IP medium. In the present study, the computed values of $D_f$ cover approximately from 0.03 to 3.4 MHz s$^{-1}$. This range fall well within the range reported by \citet{Vrsnak2002}, affirming the reliability of our analysis. Furthermore, we find a strong power-law relationship between $D_f$ and the corresponding frequency $f_L$, expressed as $D_f \propto f_L^{\sim 1.5}$. The strength of this correlation is supported by a correlation coefficient ($C$) of $\geq$ 0.75.
These findings are broadly consistent with those of \citet{Vrsnak2002}, who reported a similar power-law dependence of $D_f \propto f_L^{1.89}$ and a higher correlation coefficient ($C$ = 0.93), indicating a robust link between $D_f$ and the local plasma frequency, $f_L$. This consistency reinforces the interpretation that higher frequencies, which correspond to denser coronal regions, are associated with more rapid $D_f$ due to steeper plasma density gradients.

In contrast, \citet{Mann1995} reported significantly lower values of $D_f$, in the range of 0.04 -- 0.46 MHz s$^{-1}$, likely due to the selection of events associated with weaker shocks. This discrepancy highlights the variability in $D_f$ depending on the coronal conditions \citep[see also][]{Reiner1999, Magdalenic2012}.

\subsection{Alfv\'en Mach number}
The Alfv\'en Mach number ($M_A$) is a key parameter to describe
the strength of a shock. It is reported that the large values of $M_A$ can easily generate type II radio bursts \citep{Shen2007, Su2016}. However, \cite{Mann2022} found the $M_A$ to be in the range 1.59 -- 2.53. They concluded that this moderate $M_A$ is capable of accelerating electrons to energies high enough to excite Langmuir waves, and hence the type II radio bursts. The computed values of the $M_A$ in our study for fundamental and harmonic bands considering perpendicular and parallel shocks are in range 1.10 -- 3.38. These values of $M_A$ are consistent with those computed in previous studies \citep{Vrsnak2002, Warmuth2005, Su2016, Magdalenic2020, Mann2022}. In addition to statistical studies, the obtained here $M_A$ range is corroborated by case studies \citep{Zucca2014, Su2016}. 
The computed values of $M_A$ are always high in case of parallel shock. This could be due to the Buneman instability \citep[as described in the model of][]{Chernov2021}, according to which the shock wave can only be strictly perpendicular. If it deviates from the perpendicular by an angle of $>$ 2 $^{\circ}$, the radiation stops. At the same time, the radiation from the parallel front is much weaker \citep{Zaitsev1977}.
Additionally, for the fundamental bands, there are only two events for which the $M_A$ is greater than 2 (i.e., 2.15 and 2.47), whereas in the case of harmonic bands, the events are three (with values 2.80, 3.29, and 3.37). These deviations from the theoretical results for perpendicular shock \citep[namely values below 2, e.g.,][]{Priest1982}, could be due to the inconsistencies in the adopted visual procedure while selecting the frequency and time of the splitted bands.

\subsection{Alfv\'en speed and coronal magnetic field}
Previous studies have reported a range of Alfv\'en speeds and magnetic field strengths computed using the type II radio bursts and other phenomena. For instance, \citet{Vrsnak2002} analyzed 18 metric type II events with band splitting and found Alfv\'en speeds ranging from 400 to 500 \kms\ at a heliocentric distance of $\approx$ 2 R$_s$, increasing to 450 -- 700 \kms\ at $\approx$ 2.5 R$_s$.
\citet{Cho2007} studied the 18 August 2004 event and computed the Alfv\'en speed and magnetic field strength using type II radio burst from Green
Bank Solar Radio Burst Spectrometer (GBSRBS), associated CME from Mauna Loa Solar Observatory and, using Potential Field Source Surface (PFSS) extrapolation. They reported Alfv\'en speeds in the range of 400 -- 550 \kms\ and magnetic field strengths between 0.4 and 1.3 G, varying from a height of 1.4 to 2.1 R$_s$. 
A case study conducted by \citet{Gopalswamy2012case}, combining type II burst data and extreme-ultraviolet images, found a lower value of Alfv\'en speeds (i.e., 140 -- 460 \kms) and magnetic fields between 1.3 and 1.5 G at heights of 1.2 -- 1.5 R$_s$. Similarly, \citet{Vasanth2014} reported magnetic fields of 2.7 -- 1.7 G over 1.3 -- 1.5 R$_s$, and Alfv\'en speeds of 406 -- 600 \kms.

In comparison, the values obtained in our study show a broad range of Alfv\'en speeds and magnetic field strengths, considering both perpendicular and parallel shocks. The Alfv\'en speeds in our analysis is in the range 268 -- 1294 \kms\ and 233 -- 1177 \kms\ for fundamental and harmonic bands, respectively. The magnetic field strengths span from 0.48 -- 4.17 G for fundamental, and 0.9 -- 7.13 G for harmonic band. 

Overall, while our Alfv\'en speed and magnetic field estimates are consistent with the physical expectations from type II burst-related shock scenarios, they tend to be higher than those from earlier studies. The possible reasons could be different local coronal conditions, measurement techniques, or variations in heliocentric distance and plasma environment.
The power law fitting in the variation of magnetic field strength with the radial distances gives an estimate of the magnetic field dependence on the radial distance. This is done by \cite{Vrsnak2002} and \cite{Vrsnak2004}. It is found that the power law fit as $B\propto R^{-3.4}$ in case of former, whereas, it varies as $B\propto R^{-1.75}$ to $B\propto R^{-2.23}$ in the latter. The power-law fit of our current study fits well in this range considering both perpendicular and parallel shocks for the splitting in fundamental and harmonic bands. These values lies in the range $B\propto R^{-2.78}$ -- $B \propto R^{-2.85}$ and $B\propto R^{-2.73}$ -- $B \propto R^{-3.09}$ for fundamental and harmonic bands, respectively.

The observed discrepancies in the deduced plasma parameters from the fundamental and harmonic bands could be due to a multitude of effects, a combination of several underlying emission mechanisms or/and limitations of the applied methodology and theory.

\section{Summary}
\label{sec:summary}
Based on the 44 m-type II bursts exhibiting band-splitting analyzed in solar cycle 24 (2009 -- 2019), the following trends were identified in this study:

\begin{itemize}
\item Six out of 44 (13\%) band-splitting bursts are associated with X-class flares and only three of them continued as DH-type II.
    \item Shock speeds deduced for the fundamental band show a preference towards lower values (with the majority of cases within the range 401 -- 1000 \kms), compared to the harmonic emissions that tends to give higher shock speeds (from 601 to $>$1000 \kms).
    \item The same results for the shock heights are obtained, independent on fundamental or harmonic emission data used: a significant number of the shock waves are forming at low coronal heights (1.41 -- 1.6 $R_S$).
    \item BDW is nearly constant over the entire frequency spectrum and with the height of solar corona.
    \item Alv\'en Mach number at predominantly lower values (below 1.4) are obtained for parallel shocks, whereas for the perpendicular shocks, the Mach number distribution starts with a peak (over the range 1.21 -- 1.6) ending with a slow decline (independent on the fundamental or harmonic bands in either case).
    \item Alv\'en speed values tend to cover almost entirely the range $\approx$ 230 -- 1230 \kms in either combination of the fundamental/harmonic band with a parallel/perpendicular shock wave.
    \item Magnetic field strength for the fundamental band tends to be at lower values, in most cases below 2 G, whereas for the harmonic band it is above 2 G, independent of the shock type. 
\end{itemize}
All these results are obtained for the first time on a large time scale i.e., for the entire solar cycle 24. Some of these results confirm the previously known ones. They are a reliable aid for further theoretical research.

\appendix
\section{Details of All Events}
In the appendix, we present a table of the list of events chosen for current our study. All the events have splitting in both fundamental and harmonic bands of the dynamic radio spectra. The table also includes details about the associated flares (such as their GOES class, onset, peak, and end times), CMEs (onset, speed, acceleration, and angular width), and the DH-type II radio bursts (see Table \ref{table1}). The measured frequencies at different times from the fundamental and harmonic bands are presented in Table \ref{table2}.

\begin{table*}[!t]
\addtolength{\tabcolsep}{-1pt}
\renewcommand{\arraystretch}{1.2}
\small
    \centering
        \caption{Details of band splitted m-Type II radio bursts. The Table also presents the details of associated flares, CMEs, and DH-type II parameters. (`*' denote the associated CME on next day.)}
        \medskip
\begin{tabular}{ccccccccccc}
    \hline
Event & \multicolumn{2}{c}{m-type II Burst} & \multicolumn{3}{c}{Associated Flare} & \multicolumn{4}{c}{Associated CME} & Associated \\
\cmidrule[0.01pt](r{0.3em}){2-3}%
\cmidrule[0.01pt](r{0.3em}){4-6}%
\cmidrule[0.01pt](r{0.3em}){7-10}%

 No. & Date & Onset & Onset & GOES & Location & Onset & Speed & Acceleration & Angular width & DH-type II/ \\
 & DD-MM-YYYY & (UT) & (UT) & Class & & (UT) &(km s$^{-1}$) & (m s$^{-2}$) & (degree) & Onset (UT) \\
\hline
1 & 13-06-2010 & 05:40 & 05:30 & M1.0 & S24W91 & 06:06 & 320 & 4.8 & 16 & No \\
2  & 16-10-2010 & 19:15 & 19:07 & M2.9 & S19W29 & 20:12 &350 & 47 & 32 & No \\
3  & 12-11-2010 & 01:35 & 01:28 & C4.6 & S22W11 & 02:24 & 245 &-0.9& 61 & No \\
4  & 25-03-2011 & 23:15 & 23:08 & M1.0 & S16E30 & 01:25* &339 &-6.6& 12 & No \\
5  & 18-01-2012 & 23:20 & 22:57 & C5.1 & N24W47 & 23:48 & 270 & -2 & 119 & No \\
6  & 26-03-2012 & -- & -- & -- & Backside & 23:12 & 1390 &-32.3& 360 & 23:15 \\
7  & 24-04-2012 & 07:45 & 07:38 & C3.7 & N14E58 & 08:12 & 443 &-8.1& 192 & No \\
8  & 03-06-2012 & 18:00 & 17:48 & M3.3 & N16E33 & 18:12 & 605 &-8.7& 180 & No \\
9  & 02-07-2012 & 05:00 & 05:01 & C3.5 & S17E03 & 06:00 & 251 &-1.6& 90 & No \\
10 & 06-07-2012 & 23:10 & 23:01 & X1.1 & S17W50 & 23:24 & 1828 &-56.1& 360 & 23:10\\
11 & 12-07-2012 & 16:25 & 15:37 & X1.4 & S17W08 & 16:48 & 885 &195.6& 360 & 16:45 \\
12 & 25-09-2012 & 04:30 & 04:24 & C3.6 & N09E03 & No CME & -- & -- & -- & No \\
13 & 08-11-2012 & 02:20 & 02:08 & M1.7 & N12W71 & 02:36 & 855 &-15.2& 360 & No \\
14 & 12-11-2012 & 23:30 & 23:13 & M2.0 & S21E43 & 00:54 & 611 &3.9 & 45 & No \\
15 & 28-04-2013 & 20:15 & 19:46 & C4.4 & S16W44 & 20:48 & 497 &-23.3& 91 & No\\
16 & 02-05-2013 & 05:05 & 04:58 & M1.1 & N09W31 & 05:24 & 671 &1.1 & 99 & No\\
17 & 14-05-2013 & 01:05 & 01:00 & X3.2 & N12E67 & 01:25 & 2625 & -51 & 360 & 01:16 \\
18 & 03-07-2013 & 07:04 & 07:00 & M1.5 & S14E56 & 07:24 & 807 &-26.1& 267 & No\\
19 & 25-10-2013 & 03:00 & 02:48 & M2.9 & S08E59 & 03:24 & 344 &-3.9& 121 & No \\
20 & 25-10-2013 & 13:35 & 13:29 & C2.3 & N08W37 & No CME & -- & -- & -- & No \\
21 & 08-11-2013 & 04:25 & 04:20 & X1.1 & S11W03 & 05:12 & 409 & -5.7& 105 & No \\
22 & 08-01-2014 & 03:40 & 03:39 & M3.6 & N11W91 & 04:12 & 643 & -3.7& 108 & No \\
23 & 11-02-2014 & 03:30 & 03:22 & M1.7 & S13W00 & 04:12 & 222 & 1.4 & 81 & No \\
24 & 20-02-2014 & 03:20 & 02:48 & C3.3 & S11E32 & 03:12 & 993 &-48.8& 360 & No \\
25 & 20-02-2014 & 07:45 & 07:26 & M3.0 & S14W81 & 08:00 & 948 &-9.5 & 360 & 08:05 \\
26 & 20-03-2014 & 04:00 & 03:42 & M1.7 & S15E27 & 04:36 & 740 & -2 & 360 & No \\
27 & 01-08-2014 & 18:20 & 17:55 & M1.5 & S09E08 & 18:36 & 789 &-15.2& 360 & 18:58 \\
28 & 22-08-2014 & 00:05 & 00:03 & B8.0 & N05E13 & No CME & -- & --& -- & No \\
29 & 25-08-2014 & 15:10 & 14:46 & M2.0 & N09W47 & 15:36 & 555 & -12.2& 360 & 15:20 \\
30 & 07-09-2014 & 02:00 & 01:53 & C7.5 & S20E46 & 02:24 & 487 & -10.7& 104 & No \\
31 & 23-09-2014 & 23:10 & 23:03 & M2.3 & S11E36 & 23:36 & 331 & 18.6& 134 & 23:41 \\
32 & 28-09-2014 & 02:45 & 02:39 & M5.1 & S17W40 & 03:24 & 215 & 5.5 & 60 & No \\
33 & 02-10-2014 & 19:00 & 18:49 & M7.3 & S14W91 & 19:12 & 513 & -0.5& 159 & 21:34 \\
34 & 05-11-2014 & 09:40 & 09:26 & M7.9 & N15E53 & 10:00 & 386 & -3 & 182 & No \\
35 & 07-11-2014 & 17:20 & 16:53 & X1.6 & N15E33 & 18:08 & 795 &-24.4& 293 & No \\
36 & 17-12-2014 & 04:40 & 04:25 & M8.7 & S19W02 & 05:00 & 587 &-2.1 & 360 & 05:00 \\
37 & 20-12-2014 & 00:40 & 00:11 & X1.8 & S18W42 & 01:25 & 830 & -8.5& 257 & No \\
38 & 09-03-2015 & 23:40 & 23:29 & M5.8 & S17E39 & 00:00*& 995 &-10.2& 360 & No \\
39 & 21-12-2015 & 01:00 & 00:52 & M2.8 & N04E85 & 01:26 & 389 & -4.7& 71 & No \\
40 & 23-12-2015 & 00:30 & 00:23 & M4.7 & S23E49 & 01:25 & 520 & 1.6 & 89 & 01:18 \\
41 & 02-05-2016 & 08:40 & 08:32 & C3.5 & N21E24 & 09:12 & 262 & -3.7& 61 & No \\
42 & 04-05-2016 & 12:55 & 13:41 & C1.3 & N11W91 & 14:12 & 390 & -3.4& 134 & 14:20 \\
43 & 10-07-2016 & 01:00 & 00:53 & C8.6 & N09E49 & 01:25 & 368 & -2.5& 101 & No \\
44 & 20-10-2017 & 23:30 & 23:10 & M1.1 & S11E91 & 00:00*& 331 & -2.8& 109 & No\\
\hline
        \end{tabular}
    \label{table1}
\end{table*}

\begin{table*}[!t]
\addtolength{\tabcolsep}{-1pt}
\renewcommand{\arraystretch}{1.2}
\small
    \centering
        \caption{Band splitting parameters of the fundamental and harmonic bands.}
        \medskip
\begin{tabular}{ccccccccccccc}
    \hline
Event & \multicolumn{6}{c}{Fundamental Band} & \multicolumn{6}{c}{Harmonic Band} \\
\cmidrule[0.01pt](r{0.3em}){2-7}%
\cmidrule[0.01pt](r{0.3em}){8-13}%
Date & \multicolumn{3}{c}{Start} & \multicolumn{3}{c}{End} & \multicolumn{3}{c}{Start} & \multicolumn{3}{c}{End} \\
\cmidrule[0.01pt](r{0.3em}){2-4}%
\cmidrule[0.01pt](r{0.3em}){5-7}%
\cmidrule[0.01pt](r{0.3em}){8-10}%
\cmidrule[0.01pt](r{0.3em}){11-13}%
 & UFB & LFB & Time & UFB & LFB & Time & UFB & LFB & Time & UFB & LFB & Time \\ 
(DD-MM-YYYY) & (MHz) & (MHz) & (UT) & (MHz) & (MHz) & (UT) & (MHz) & (MHz) & (UT) & (MHz) & (MHz) & (UT) \\
\hline

13-06-2010 & 82 & 62 & 05:40:35 & 46 & 38 & 05:45:02 & 171 & 137 & 05:39:36 & 59 & 49 & 05:47:50 \\
16-10-2010 & 133 & 119 & 19:13:32 & 82 & 56 & 19:16:40 & 178 & 148 & 19:14:17 & 120 & 83 & 19:17:26 \\
12-11-2010 & 91 & 73 & 01:40:46 & 60 & 45 & 01:45:50 & 152 & 134 & 01:40:46 & 98 & 92 & 01:45:46 \\
25-03-2011 & 111 & 93 & 23:17:30 & 45 & 38 & 23:24:37 & 175 & 151 & 23:18:44 & 85 & 70 & 23:25:02 \\
18-01-2012 & 38 & 33 & 23:24:50 & 30 & 26 & 23:28:08 & 68 &  57 & 23:26:10 & 51 & 44 & 23:30:00 \\
26-03-2012 & 88 & 75 & 22:50:10 & 53 & 30 & 22:57:50 & 164 & 130 & 22:50:35 & 96 & 60 & 23:01:10 \\
24-04-2012 & 45 & 40 & 07:48:24 & 39 & 26 & 07:51:43 & 115 & 78 & 07:48:40 & 45 & 40 & 07:54:40 \\
03-06-2012 & 68 & 54 & 18:01:09 & 34 & 27 & 18:09:30 & 131 & 85 & 18:00:30 & 52 & 30 & 18:14:26 \\ 
02-07-2012 & 95 & 84 & 05:09:47 & 62 & 55 & 05:12:08 & 166 & 150 & 05:09:53 & 87 & 81 & 05:14:08 \\
06-07-2012 & 44 & 33 & 23:11:20 & 32 & 26 & 23:13:05 & 104 & 71 & 23:10:26 & 50 & 34 & 23:15:00 \\
12-07-2012 & 44 & 37 & 16:27:48 & 34 & 28 & 16:34:27 & 82 & 58 & 16:27:57 & 47 & 36 & 16:37:21 \\
25-09-2012 & 80 & 76 & 04:29:43 & 54 & 50 & 04:32:16 & 160 & 137 & 04:29:07 & 106 & 91 & 04:33:46 \\
08-11-2012 & 64 & 52 & 02:22:02 & 39 & 27 & 02:28:35 & 120 & 92 & 02:21:50 & 49 & 39 & 02:37:56 \\
12-11-2012 & 78 & 62 & 23:29:30 & 53 & 45 & 23:35:20 & 159 & 126 & 23:29:30 & 96 & 80 & 23:34:05 \\
28-04-2013 & 51 & 36 & 20:19:45 & 34 & 27 & 20:23:35 & 83 & 68 & 20:19:47 & 45 & 38 & 20:29:44 \\
02-05-2013 & 47 & 36 & 05:08:24 & 30 & 25 & 05:13:00 & 154 & 83 & 05:07:15 & 70 & 45 & 05:14:50 \\
14-05-2013 & 84 & 60 & 01:09:35 & 52 & 30 & 01:17:45 & 180 & 147 & 01:08:35 & 102 & 61 & 01:17:10 \\
03-07-2013 & 58 & 40 & 07:13:20 & 32 & 25 & 07:17:29 & 132 & 104 & 07:14:00 & 84 & 59 & 07:17:56 \\
25-10-2013 & 57 & 44 & 03:00:38 & 37 & 29 & 03:06:23 & 115 & 82 & 03:00:26 & 67 & 50 & 03:07:47 \\
25-10-2013 & 83 & 74 & 13:37:05 & 68 & 61 & 13:38:47 & 152 & 139 & 13:37:14 & 128 & 113 & 13:38:50 \\
08-11-2013 & 120 & 72 & 04:28:42 & 41 & 34 & 04:35:06 & 174 & 128 & 04:29:27 & 83 & 70 & 04:34:57 \\
08-01-2014 & 84 & 76 & 03:51:12 & 43 & 37 & 03:56:21 & 174 & 145 & 03:50:40 & 88 & 77 & 03:55:45 \\
11-02-2014 & 55 & 43 & 03:32:18 & 32 & 26 & 03:35:54 & 131 & 102 & 03:31:57 & 56 & 42 & 03:40:20 \\
20-02-2014 & 60 & 52 & 03:22:37 & 45 & 40 & 03:25:37 & 118 & 102 & 03:22:34 & 93 & 79 & 03:25:25 \\
20-02-2014 & 86 & 73 & 07:45:31 & 53 & 37 & 07:49:50 & 169 & 132 & 07:45:34 & 83 & 56 & 07:52:25 \\
20-03-2014 & 48 & 39 & 04:01:18 & 35 & 31 & 04:05:18 & 98 & 80 & 04:00:50 & 69 & 58 & 04:06:00 \\
01-08-2014 & 42 & 35 & 18:18:58 & 32 & 26 & 18:24:28 & 70 & 72 & 18:18:58 & 35 & 29 & 18:35:10 \\
22-08-2014 & 121 & 102 & 00:09:14 & 91 & 78 & 00:10:44 & 176 & 146 & 00:09:50 & 102 & 92 & 00:13:40 \\
25-08-2014 & 50 & 43 & 15:09:08 & 32 & 26 & 15:14:02 & 83 & 73 & 15:10:23 & 56 & 26 & 15:23:08 \\
07-09-2014 & 82 & 70 & 02:04:20 & 62 & 52 & 02:06:25 & 180 & 151 & 02:03:10 & 84 & 73 & 02:08:50 \\
23-09-2014 & 76 & 66 & 23:17:45 & 38 & 33 & 23:26:18 & 175 & 156 & 23:15:27 & 109 & 83 & 23:20:42 \\
28-09-2014 & 91 & 84 & 02:46:02 & 50 & 45 & 02:53:09 & 171  &138 & 02:46:20 & 97 & 85 & 02:53:42 \\
02-10-2014 & 59 & 47 & 19:03:00 & 48 & 41 & 19:05:55 & 146 & 112 & 19:01:10 & 81 & 55 & 19:10:05 \\
05-11-2014 & 167 & 126 & 09:43:56 & 52 & 42 & 09:52:00 & 175 & 137 & 09:46:44 & 75 & 60 & 09:56:17 \\
07-11-2014 & 85 & 64 & 17:20:40 & 36  &30 & 17:28:50 & 177 & 145 & 17:20:00 & 75 & 59 & 17:28:50 \\
17-12-2014 & 42 & 32 & 04:44:00 & 31 & 25 & 04:48:15 & 67 & 52 & 04:46:33 & 36 & 26 & 05:01:12 \\
20-12-2014 & 44 & 35 & 00:47:50 & 32 & 26 & 00:52:14 & 88 & 70 & 00:47:44 & 36 & 29 & 01:04:23 \\
09-03-2015 & 92 & 77 & 23:42:10 & 51 & 43 & 23:52:55 & 177 & 140 & 23:42:35 & 103 & 78 & 23:52:40 \\
21-12-2015 & 79 & 65 & 01:01:49  &59 & 47 & 01:03:58 & 162  &128 & 01:01:46 & 35 & 28 & 01:21:25 \\
23-12-2015 & 67 & 56 & 00:36:10 & 31 & 25 & 00:41:45 & 130  &98 & 00:36:20 & 36 & 26 & 00:48:39 \\
02-05-2016 & 49 & 38 & 08:41:06 & 33 & 25 & 08:43:27 & 94 & 62 & 08:41:15 & 42 & 32 & 08:49:00 \\
04-05-2016 & 57 & 43 & 13:55:22 & 41 & 33 & 13:58:19 & 83 & 71 & 13:57:25 & 41 & 35 & 14:09:30 \\
10-07-2016 & 82 & 65 & 01:01:07 & 44 & 32 & 01:04:25 & 172 & 124 & 01:00:28 & 79 & 60 & 01:05:19 \\
20-10-2017 & 88 & 70 & 23:30:05 & 41 & 33 & 23:38:30 & 172 & 127 & 23:30:10 & 84 & 62 & 23:37:20 \\
\hline
        \end{tabular}
    \label{table2}
\end{table*}

\section*{Acknowledgments}
The authors thank Dr. G. P. Chernov and the other anonymous reviewer for their valuable comments and suggestions on the manuscript.
We thank the open data policy of RSTN, GOES, and SOHO for their open data policy. RC acknowledges the support from the DST/SERB project number EEQ/2023/000214.  
PD acknowledges the support by the Research Council of Norway through its Centres of Excellence scheme, project number 262622.

\bibliographystyle{jasr-model5-names}
\biboptions{authoryear}
\bibliography{refs}

\end{document}